\documentclass[aps,floatfix,preprint,nofootinbib]{revtex4}
\usepackage[parfill]{parskip}    
\usepackage{graphicx}
\usepackage{amssymb}
\usepackage{epstopdf}
\usepackage{verbatim}
\usepackage{url}
\newcommand{\be}{\begin{equation}}
\newcommand{\bea}{\begin{eqnarray}}
\newcommand{\beq}[1]{\begin{equation}\label{#1}}

\newcommand{\ee}{\end{equation}}
\newcommand{\eea}{\end{eqnarray}}
\newcommand{\eeq}{\end{equation}}
\newcommand{\lsim}{\!\mathrel{\hbox{\rlap{\lower.55ex \hbox{$\sim$}} \kern-.34em \raise.4ex \hbox{$<$}}}}
\newcommand{\gsim}{\!\mathrel{\hbox{\rlap{\lower.55ex \hbox{$\sim$}} \kern-.34em \raise.4ex \hbox{$>$}}}}
\newcommand{\mr}[1]{\mathrm{#1}}

\begin{document}

\setlength{\baselineskip}{0.22in}

\begin{flushright}MCTP-10-01 \\
\end{flushright}
\vspace{0.2cm}

\title{On the Correlation Between the Spin-Independent and Spin-Dependent Direct Detection of Dark Matter}

\author{Timothy Cohen, Daniel J. Phalen and Aaron Pierce}
\vspace{0.2cm}
\affiliation{Michigan Center for Theoretical Physics (MCTP) \\
Department of Physics, University of Michigan, Ann Arbor, MI
48109}

\date{\today}

\begin{abstract}
We study the correlation between spin-independent and spin-dependent scattering in the context of MSSM neutralino dark matter for both thermal and non-thermal histories.  We explore the generality of this relationship with reference to other models. We discuss why either fine-tuning or numerical coincidences are necessary for the correlation to break down.  We derive upper bounds on spin-dependent scattering mediated by a $Z^0$ boson.
\end{abstract}

\maketitle

\section{Introduction}
There is compelling evidence for dark matter (DM) from  a variety of astrophysical  observations:   rotation rates of galaxies, gravitational lensing, and temperature perturbations of the cosmic microwave background (CMB).  The WMAP experiment, which measures the CMB, provides   the best value for the DM relic density \cite{Komatsu:2008hk},
\be\label{eq:OmegaDMWMAP}
\Omega_\mr{DM}\,h^2 = 0.1131\pm 0.0034.
\ee
While there are potential hints of DM in cosmic ray anomalies (\emph{e.g.} PAMELA \cite{PAMELA} and Fermi \cite{FERMI}) and from the annual modulation signal seen by the DAMA/LIBRA collaboration \cite{DAMA, DAMALIBRA}, there is currently no measurement  which gives conclusive information about the DM's properties.   The identity of the particle responsible for $\Omega_\mr{DM}$ remains a mystery.

One DM candidate worthy of special attention is the Weakly Interacting Massive Particle (WIMP). The strongest motivation is the so called ``WIMP miracle"  -- a thermally produced, stable particle with a weak scale mass and perturbative couplings will freeze-out with a relic density of the right order of magnitude to constitute the DM \cite{Lee:1977ua}.  When  added to the expectation of new physics at $\mathcal{O}$(TeV) responsible for the stabilization of the weak scale, the WIMP paradigm becomes even more compelling: the new physics often includes novel symmetries, which result in at least one particle being stable on cosmological timescales.  If this particle is weakly interacting then it can be WIMP DM.

However, this picture turns out to be an over-simplification.  Not just any weak scale stable particle will do.  If the DM is weakly interacting in the strictest sense -- \emph{i.e.} has full-strength $SU(2)_L \times U(1)_{Y}$ gauge interactions -- then DM may be excluded by existing direct detection (DD) experiments.  In particular, a weak-scale Dirac (vector-like) fermion, $\chi_\mr{D}$, with $SU(2)$ interactions (which encompasses the simplest DM model of all, a Dirac neutrino), feels the weak force via the operator:
\begin{equation}
\mathcal{O}_\mr{vector} = (\bar{\chi}_\mr{D}\, \gamma^{\mu}\, \chi_\mr{D})\,Z^0_{\mu}. \label{eqn:Zvector}
\end{equation}
When the coefficient of this operator is typical in size, namely $\mathcal{O}(g/ \cos{\theta_{w}})$, where $g$ is the $SU(2)$ coupling constant and  $\theta_w$ is the weak mixing angle, it leads to a huge DD signal -- experiments constrain the DM mass to be greater than 50 TeV \cite{Servant:2002hb}.  Furthermore, the thermal relic density for a 50 TeV Dirac neutrino will be far too large to explain the WMAP measurement.  Thus, DM at the weak scale requires a strong suppression of this operator.  In fact, it is straightforward to eliminate it entirely.  If $\chi$ is a Majorana spinor, the operator $(\bar{\chi}\,\gamma^{\mu}\,\chi)\,Z^0_{\mu}$ identically vanishes due to the properties of Majorana bilinears.    The DM may be Majorana if an $SU(2)$ singlet Majorana fermion mixes with a Dirac state.  This mixing can only be accomplished via $SU(2)$ breaking in the WIMP sector, \emph{i.e.} through a Higgs boson vacuum expectation value (vev).  Then the resultant DM particle has a non-zero coupling to a Higgs boson, $h$, and the dominant scattering process is due to the following operators:
\bea
\mathcal{O}_\mr{Higgs} &=&  (\bar{\chi} \,\chi)\,h, \label{eq:OpHiggsChiChi}\\
\mathcal{O}_{Z^0} &=& (\bar{\chi}\,\gamma^{\mu}\gamma^5\,\chi)\,Z^0_{\mu}.\label{eq:OpZChiChi}
\eea
In a multi-Higgs boson theory, $h$ need not be {\it the} Higgs boson of the Standard Model (SM), but even in these theories, there often is a Higgs boson that has SM-like properties.  We will explore the impact of these operators on Spin-Independent (SI) and Spin-Dependent (SD) scattering off of nuclei, paying particular attention to the expected correlation between the rates at these two types of experiments.

While we perform most of our analysis in the context of the Minimal Supersymmetric Standard Model (MSSM) (for a review of the MSSM, see \cite{Martin:1997ns}), we reference other models where appropriate to emphasize the generality of our arguments.  We will review the assertion that post-LEP (largely due to the constraints on the chargino and slepton masses), one may consider a mixed or ``well-tempered" neutralino as a likely DM candidate, if it is thermally produced \cite{ArkaniHamed:2006mb}.  We will show that in this case, light Higgs boson and $Z^0$ exchange will generically lead to a signal in the next generation of SI and SD experiments.

A thermal history for the WIMP is not the only possibility.  For example, non-thermal mechanisms may populate the DM (\emph{e.g.} through the decay of a modulus or gravitino \cite{Moroi:1999zb}), or the DM can be overabundant and subsequently diluted by extra sources of entropy.  These options allow a WIMP with a wider range of properties, since the annihilation rate is not fixed by the thermal history.  In what follows, we do not rescale DD signals to the (too-low/too-high) thermal relic density.    \emph{In all cases, we assume that the $\mr{WIMP}$ constitutes the total $\mr{DM}$ density, determined from astrophysical measurements to be $\rho_\mr{DM} \approx 0.3\,\mr{GeV/cm}^{3}$.}  We will be clear when we are making the assumption of a thermal history.  For the purposes of this study, a ``thermal" WIMP is one whose thermal relic density is within the generous range $\pm 3\,\sigma$ of the WMAP measurement given in Eq.~(\ref{eq:OmegaDMWMAP}).  We also note that more recent determinations favor a slightly larger value: $\rho_\mr{DM} = 0.39-0.43$ GeV/cm$^3$ \cite{Catena:2009mf, Salucci:2010qr}.  This would extend the reach of the direct detection experiments by a factor of $\approx 4/3$ and probe more parameter space.  An accurate determination of the local DM density is important for an accurate measurement of the DM DD cross section.

Related results already exist in the literature, including some comprehensive numerical scans.  However, we find that often the (simple) underlying physics is left obscure.  We hope to make clear the expected size of various contributions to DD and the relationship to the assumption of a  thermal relic abundance.  Assuming there are no conspiratorial cancellations, these typical sizes represent important targets for DD experiments.

There is an overwhelming literature in existence on the subject of  DD, see reviews \cite{Gaitskell:2004gd,Jungman:1995df} and references therein.  Of particular interest to us is the relationship between the size of the SD and SI signals, which has recently been explored in \cite{Barger:2008qd, Belanger:2008gy, Bertone:2007xj, Berger:2008cq}.

In the next section, we begin by discussing the current experimental status and then make naive estimates for the SI and SD DD cross sections from $h$ and $Z^0$ exchange respectively.  In Sec. \ref{sec:NeutralinoDM} we lay out the specific structure of the SI and SD operators in the MSSM and estimate the naive size of the SI and SD cross sections.  Then in Sec. \ref{sec:ArgueWellTempered} we review the argument for a well-tempered neutralino and discuss some alternatives.  Sec. \ref{sec:SDphys} concentrates on illuminating the expected size of the SD cross section for mixed DM models with various restrictions.   In Sec.  \ref{sec:SIvsSD} we describe the conditions under which SI and SD signals in the MSSM are expected to be correlated.  Technical results are relegated to three appendices.

\section{Direct Detection Preliminaries}\label{sec:DD}
\begin{figure}[h]
\begin{center}
\includegraphics[width=1.0\textwidth]{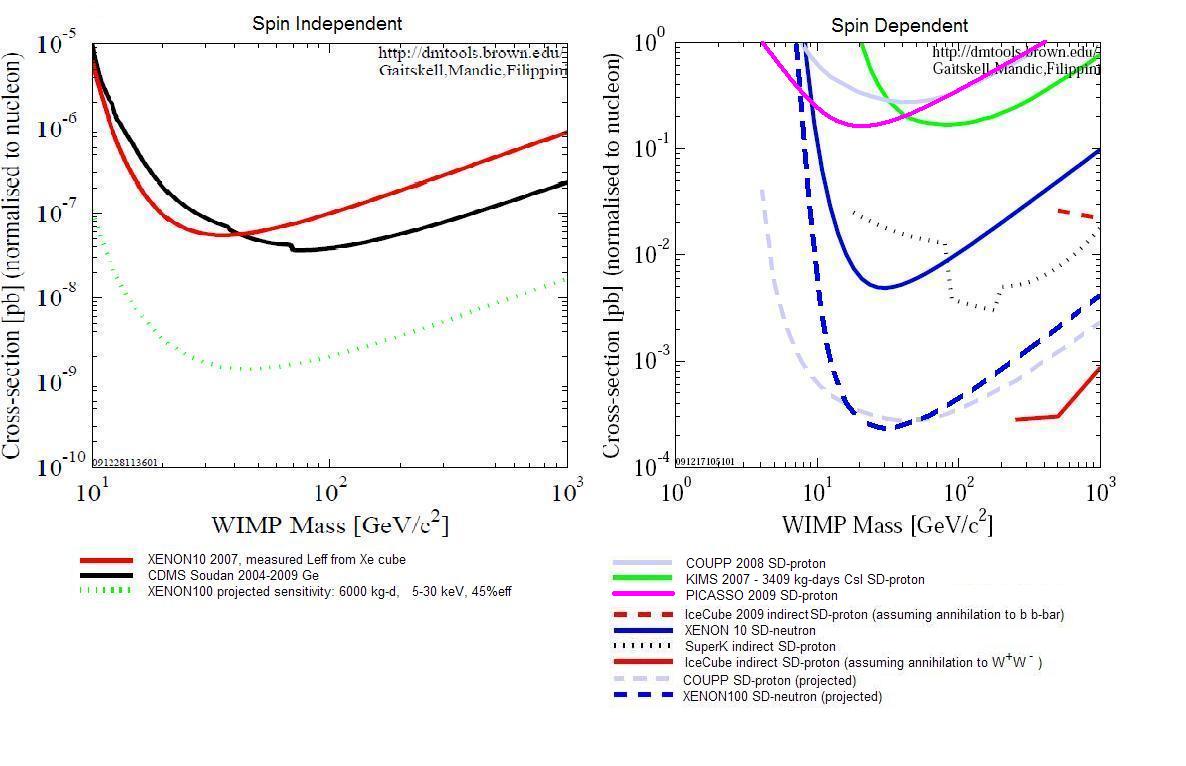}
\end{center}
\caption{Current bounds on SI (left) and SD (right) DM-nucleon cross sections.  The COUPP and XENON100 projected SD bounds are only estimates -- we have scaled the current exclusion curve of COUPP by a factor of $10^{-3}$ \cite{LevineConversation} and the current SD exclusion curve of XENON10 by the factor which scales the XENON10 SI limit to the XENON100 SI limit.}
\label{fig:Bounds}
\end{figure}

The interactions in Eqs.~(\ref{eq:OpHiggsChiChi}) and (\ref{eq:OpZChiChi}) lead to SI and SD elastic signals in DD experiments, respectively.  In Fig. \ref{fig:Bounds} we have plotted the current experimental limits for SI and SD DD.  Currently, the state of the art SI experiments are CDMS \cite{CDMS} and XENON \cite{XENON}.  XENON constrains $\sigma^{\chi\,p}_\mr{SI} < 4.5\times 10^{-8}$ pb for $m_{\chi} = 30$ GeV.  After combining their most recent run with previous data, CDMS-II has a 90\% CL bound of 3.8 $\times$ 10$^{-8}$ pb for a WIMP with a mass of 70 GeV \cite{Ahmed:2009zw}.  In the most recent data set, two tantalizing events were seen, but it is premature to attribute these to signal.  In any case, XENON100 expects to place a limit on the order of $\sigma^{\chi\,p}_\mr{SI} \approx$ few $\times 10^{-9}$ pb by early 2010.  Thus, we will consider SI cross sections greater than $5\times 10^{-9}$ pb as potentially probeable in the short-term, and hence ``large."

There are two ways the SD cross sections are constrained.  The first is via DD experiments. The current best bound on the SD DM-proton interaction comes from the KIMS experiment \cite{KIMS}, $\sigma^{\chi\,p}_\mr{SD} < 1.6\times 10^{-1}$ pb for $m_{\chi} = 70$ GeV; the best bound on the SD DM-neutron interaction coming from the XENON experiment, $\sigma^{\chi\,n}_\mr{SD} < 6\times 10^{-3}$ pb for $m_{\chi} = 20$ GeV, with the strongest bounds for masses of ${\mathcal O}$(10) GeV coming from PICASSO \cite{Archambault:2009sm}.  There are also bounds from DM capture in the sun, assuming (as is the case in the MSSM) that the DM has annihilation products which give rise to relatively hard neutrinos.  Assuming annihilation of the DM to $W^{\pm}$ bosons is appreciable (as is  appropriate for much of the parameter space considered here, see Sec.~\ref{sec:ArgueWellTempered}), IceCube \cite{Abbasi:2009uz} places very strong bounds for masses above 250 GeV with the strongest bounds coming at 250 GeV, $\sigma^{\chi\,p}_\mr{SD} < 3 \times 10^{-4}$ pb.  At present, no limits exist from IceCube below this mass.  For smaller masses, the best limits of this type come from SuperK \cite{Desai:2004pq},  $\sigma^{\chi\,p}_\mr{SD} < 10^{-2}$ pb above $m_\chi > 20$ GeV.   

Perhaps within the next two years \cite{DPFTalk}, the COUPP \cite{COUPP} and PICASSO \cite{PICASSO} experiments will take data with a projected sensitivity to SD scattering of $\sigma^{\chi\,p}_\mr{SD} \approx 10^{-4}$ pb.  They will also have sensitivity down to much lower masses than the neutrino experiments.  The XENON data will probe $\sigma^{\chi\,p}_\mr{SD} \approx 4\times 10^{-3}$ pb for a 30 GeV WIMP.  A 1 ton COUPP-like proposed experiment \cite{Bertone:2007xj}, might ultimately probe values as low as  $10^{-7}$ pb.  The DeepCore extension to the IceCube detector should be able to extend down to the $10^{-5}$ pb level with 5 years of data \cite{Wiebusch:2009jf}.  Bounds from neutrino experiments can be avoided if particular final states dominate WIMP annihilation, \emph{e.g.} 1$^\mr{st}$ generation quarks, though this does not happen in the MSSM.  We consider SD cross sections greater than $10^{-4}$ pb as potentially achievable in the short-term, and hence ``large."  

\subsection{Spin Independent}\label{sec:SI}
The operator responsible for SI DM-nucleus interactions is 
\be\label{eq:SIOpp}
\mathcal{O}^\mr{SI}_q = c_q\, (\bar{\chi}\, \chi)\,(\bar{q}\, q),
\ee
where $\chi$ is the DM and $q$ is a quark.  Taking the expectation value of this operator between two nucleon states ($N = p$ (proton) or $n$ (neutron)) determines the effective interaction of the DM with a nucleon,
\be
\langle N | m_q\, \bar{q}\, q | N \rangle = m_N\, f^{(N)}_{Tq},
\ee
where the nuclear matrix element $f^{(N)}_{Tq}$ is determined in chiral perturbation theory from the pion nucleon-scattering sigma term. The coefficient of the effective DM-nucleon interaction, $f_N (\bar{\chi}\, \chi)(\bar{N}\, N)$, is given by
\be
\frac{f_N}{m_N} = \sum_{q=u,d,s} f_{Tq}^{(N)}\,\frac{1}{m_q}\, c_q +\frac{2}{27}\,f^{(N)}_{TG}\sum_{q=c,b,t}\frac{1}{m_q}\,c_{q}^{(h)},
\ee
where $f_{TG}^{(N)} = 1 -\sum_{q=u,d,s} f_{Tq}^{(N)}$ and the $h$ on $c_{q}^{(h)}$ refers to Higgs boson exchange \cite{Shifman:1978zn}.

The nucleon-Higgs interaction is coherent over the nucleus \cite{Freedman:1977xn} resulting in the well known $A^2$ enhancement for SI cross sections.  To compare between experiments using different nuclei, the elastic scattering cross section is normalized to a per nucleon value \cite{Gaitskell:2004gd}:
\be
\sigma_\mr{SI}(\chi\,N\rightarrow \chi\,N) = \frac{4}{\pi}\, m_r^2\,\frac{1}{A^2}\, (Z\,f_p + (A-Z)\,f_n)^2,
\ee
where $m_r$ is the reduced mass between the DM and a nucleon.  

We use the DarkSUSY package for numerical analysis \cite{DarkSUSY}, so for analytic estimates we will use the same values for the nuclear matrix elements, namely
\[\begin{array}{cccc}
f_{Tu}^{(p)} = 0.023 & f_{Td}^{(p)} = 0.034 & f_{Ts}^{(p)} = 0.14 & f_{TG}^{(p)} = 0.803\\
f_{Tu}^{(n)} = 0.019 & f_{Td}^{(n)} = 0.041 & f_{Ts}^{(n)} = 0.14 & f_{TG}^{(n)} = 0.800.
\end{array} \]
Since these values are derived from the pion nucleon scattering sigma term, their error bars are correlated.    

If the Higgs boson, $h$, that mediates the interaction between the DM and the nucleon is SM-like, the coefficients $c_q$ are given by
\be\label{eq:naievecq}
c_q = y_q\,y_\chi\, \frac{1}{m_h^2}, 
\ee
where $y_q$ ($y_\chi$) is the Yukawa coupling for the quark (DM) and $m_h$ is the Higgs mass.  The per nucleon cross section is then
\be\label{eq:EstimateSI}
\sigma_\mr{SI}(\chi\,N\rightarrow \chi\,N) \approx  5\times10^{-8} \textrm{ pb} \,  \left(\frac{y_{\chi}}{0.1}\right)^2 \left(\frac{115 \textrm{ GeV}}{m_h} \right)^4 \qquad \textrm{(SI typical)}.
\ee

Estimates based on recent lattice simulations seem to favor smaller values for the nuclear matrix elements \cite{Giedt:2009mr}.  If these lattice results are correct, the dominant contribution to the SI scattering cross section would be due to the heavy quark content of the nucleon (since $f_{TG}^{(N)}\rightarrow 1$ in the limit of small $f_q^{(N)}$) and the coefficient in Eq.~(\ref{eq:EstimateSI}) would be replaced by $2\times 10^{-8}$ pb.  In cases where $c_d \gg c_u$, which can occur in models with multiple Higgs bosons such as the MSSM, then uncertainties in the $f_{Tq}^{(N)}$ can lead to as much as an order of magnitude variation in $\sigma_\mr{SI}(\chi\,N\rightarrow \chi\,N)$ \cite{EllisOliveSavage}.

\subsection{Spin Dependent}\label{sec:SD}
The operator responsible for SD DM-nucleus interactions is 
\be\label{eq:SDOpp}
\mathcal{O}^\mr{SD}_q = d_q\, (\bar{\chi}\, \gamma^\mu \gamma^5\, \chi)(\bar{q}\, \gamma_\mu \gamma^5\, q).
\ee
Taking the expectation value of this operator between two nucleon states allows us to find the effective SD interaction of the DM with a nucleon ($N = p$ (proton) or $n$ (neutron)),
\be
\langle N | \bar{q}\, \gamma_\mu \gamma^5\, q | N \rangle = 2\, s_\mu^{(N)}\, \Delta q^{(N)},
\ee
where $s_\mu^{(N)}$ is the spin of the nucleon and the $\Delta q^{(N)}$ are extracted from polarized deep elastic scattering.  The coefficient of the effective DM-nucleon interaction, $2\,a_N(\bar{\chi}\, \gamma^\mu \gamma^5\, \chi)(\bar{N}\, s^{(N)}_\mu\, N)$, is given by
\be
a_N = \sum_{q=u,d,s} d_q\, \Delta q^{(N)}.
\ee
The elastic scattering cross section quoted by the experiments is between the DM and a nucleon which is given by
\be
\sigma_\mr{SD}(\chi\,N\rightarrow \chi\,N) = \frac{6}{\pi}\, m_r^2\, a_N^2,
\ee
where $m_r$ is the reduced mass between the DM and a nucleon.

Again we follow DarkSUSY and use the following values for the SD calculations, 
\bea
&\Delta_{u}^{(p)} = 0.77 \,\,\,\,\,\,\,\, \Delta_{d}^{(p)} = -0.40 \,\,\,\, \Delta_{s}^{(p)} = -0.12\,& \nonumber \\
&\Delta_{u}^{(n)} = -0.40 \,\,\,\,\Delta_{d}^{(n)} = 0.77 \,\,\,\,\,\, \Delta_{s}^{(p)} = -0.12.&
\eea
The prediction for SD scattering is somewhat more robust to variation in the hadronic matrix elements than the SI case:  the uncertainties in these values can lead to $\mathcal{O}(30\%)$ variation in the SD cross section \cite{EllisOliveSavage}.

If the SD interaction is mediated by the $Z^0$ boson, then the coefficients $d_q$ are given by  
\be\label{eq:EstimateSD}
d_q =\frac{g^2}{2\, c_w^2} T_3^q\, \left(\frac{Q_{Z-\mr{DM}}}{2}\right) \frac{1}{m_Z^2}, 
\ee
where $Q_{Z-\mr{DM}}$ parametrizes the coupling of the DM to the $Z^0$ and $c_w \equiv \cos{\theta_w}$.  For concreteness, (and since it is relevant for calculations of solar capture) when we quote values for SD scattering we will focus on the cross section off of protons.  For SD scattering mediated by the $Z^0$, the neutron scattering is $\mathcal{O}(20\%)$ smaller.  The SD cross section is
\be
\sigma_\mr{SD}(\chi\,p\rightarrow \chi\,p) \approx  4\times10^{-4}\, \mr{pb}\,\left(\frac{Q_{Z-\mr{DM}}}{0.1}\right)^2 \qquad \textrm{(SD typical)}.
\ee

In the next section, we discuss the form that $Q_{Z-\mr{DM}}$ takes in the MSSM.  

\section{Direct Detection of Neutralino dark matter}\label{sec:NeutralinoDM}
The best DM candidate in the MSSM is the lightest neutralino, which is an admixture of Bino ($\tilde{B}$), Wino ($\tilde{W}$), and the up and down-type Higgsinos ($\tilde{H}_u$ and $\tilde{H}_d$).  The stability of the lightest superpartner (LSP) is guaranteed by $R$-parity, which is introduced to avoid proton decay.  The neutralino mass matrix is given by 
\[ \mathcal{M}=\left( \begin{array}{cccc}
M_1 & 0   & -m_Z\,s_w \, c_{\beta} & m_Z\,s_w\, s_{\beta}\\
0   & M_2 & m_Z\,c_w\, c_{\beta} & -m_Z\,c_w\, s_{\beta}\\
-m_Z\,s_w\, c_{\beta} & m_Z\,c_w\, c_{\beta} & 0 & -\mu  \\
 m_Z\,s_w\, s_{\beta} & -m_Z\,c_w\, s_{\beta} & -\mu & 0  \end{array} \right),\]

where $M_1$ is the Bino mass, $M_2$ is the Wino mass, $\mu$ is the Supersymmetric (SUSY) Higgs boson mass parameter, $m_Z$ is the $Z^0$ mass, $\beta = \mr{arctan}(v_u/v_d)$, $v_{u,d}$ are the up and down-type Higgs boson vevs, $s_w \equiv \sin{\theta_w}$, $c_w \equiv \cos{\theta_w}$, $s_{\beta} \equiv \sin{\beta}$, and $c_{\beta} \equiv \cos{\beta}$.  

The composition of the lightest neutralino, which we denote $\chi$, is specified by 
\be
\chi \equiv Z_B\,\tilde{B}+Z_W\,\tilde{W}\,+Z_{H_d}\,\tilde{H}_d+Z_{H_u}\,\tilde{H}_u.
\ee

If squarks are heavy, the only potentially sizable contributions to SI DD are from both CP-even Higgs bosons, $h$ and $H$, where $m_h < m_H$.  We comment on the typically subdominant squark exchange contributions in Appendix \ref{sec:SquarkExchange}.  The Higgs boson exchange contributions are \cite{Barbieri:1988zs, Ellis:2000jd},
\bea
\frac{c_u}{m_u} &=& - \frac{g^2 (Z_W-t_w\,Z_B)}{4\,m_W\,s_{\beta}}\left[ \left(Z_{H_d}s_{\alpha}\,c_{\alpha} + Z_{H_u} \,c_{\alpha}^2 \right)\frac{1}{m_h^2}+\left(-Z_{H_d}s_{\alpha}\, c_{\alpha} + Z_{H_u}\, s_{\alpha}^2 \right)\frac{1}{m_H^2}\right] \label{eq:cuMSSM}\\
\frac{c_d}{m_d} &=& \,\,\,\,\,\frac{g^2 (Z_W-t_w\,Z_B)}{4\,m_W\,c_{\beta}} \left[ \left(Z_{H_u}s_{\alpha}\,c_{\alpha} + Z_{H_d}\, s_{\alpha}^2 \right)\frac{1}{m_h^2}+\left(-Z_{H_u} s_{\alpha}\, c_{\alpha} + Z_{H_d}\, c_{\alpha}^2 \right)\frac{1}{m_H^2}\right],\label{eq:cdMSSM}
\eea
where $c_{u,d}$ are the SI operator coefficients given in Eq.~(\ref{eq:SIOpp}), $g$ is the $SU(2)$ gauge coupling, $m_W$ is the $W^{\pm}$ mass, $t_w \equiv \tan{\theta_w}$, $\alpha$ is the Higgs mixing angle, $c_{\alpha} \equiv \cos{\alpha}$ and $s_{\alpha} \equiv \sin{\alpha}$.  In the decoupling ($m_H \rightarrow \infty$ and $\alpha\rightarrow \pi/2+ \beta$) and large $t_{\beta}$ limits, these expressions simplify:
\bea
\frac{c_u}{m_u} &=& \frac{- g^2}{4\,m_W}\,(Z_W-t_{w}\,Z_B)\,\frac{s_{\beta}}{m_h^2}\,Z_{H_u}, \label{eq:cuDecoupling}\\
\frac{c_d}{m_d} &=& \frac{c_u}{m_u} \left(1-\frac{t_{\beta}}{s_{\beta}^2} \frac{m_h^2}{m_H^2} \frac{Z_{H_d}}{Z_{H_u}}\right),\label{eq:cdDecoupling}
\eea
where we have only kept the $t_{\beta}$ enhanced contribution from $H$.  We will use these expressions below in Sec. \ref{sec:SIvsSD} when analyzing the allowed suppression of the SI cross section. 

The lack of an observation of a Higgs boson at LEP makes it likely that we live in at least a moderate $t_{\beta}$ regime (so that the tree-level contribution to the Higgs boson mass  $m_{h} = m_Z \cos{ 2 \beta}$  is maximized), and constraints on the mass of the charged Higgs from flavor experiments point to the decoupling limit.  Therefore, Eqs.~(\ref{eq:cuDecoupling}) and (\ref{eq:cdDecoupling}) are particularly useful for estimating the expected size of scattering.  In Sec. \ref{sec:ArgueWellTempered} we will argue for the typical size of the various neutralino mixing angles which lead to SI cross sections of the order,
\bea\label{eq:SIinMSSM}
&&\sigma_\mr{SI}^\mr{MSSM}(\chi\,N\rightarrow \chi\,N) \approx \nonumber \\
&&5 \times 10^{-9}\, \mr{pb}\,\left(\frac{\textrm{115 GeV}}{m_h}\right)^4 \left(\frac{(Z_W-t_{w}\,Z_B)\, Z_{H_u}}{0.1}\right)^2 \qquad \textrm{(MSSM: SI typical)},
\eea
where we have used Eqs.~(\ref{eq:cuDecoupling}) and (\ref{eq:cdDecoupling}) and taken $m_H \rightarrow \infty$.

In the heavy squark limit, contributions to SD DD come from $Z^0$ exchange.  Since the Bino and Wino are both $SU(2)$ singlets, they do not couple to the $Z^0$.  Therefore, SD is controlled by the Higgsino content of the WIMP.  The $Z^{0}$ exchange contribution takes the form:
\be
d_q = - \frac{g^2}{4\,m_Z^2\,c_w^2}\left(|Z_{H_d}|^2-|Z_{H_u}|^2\right)T_3^q.
\ee
A non-zero Higgsino component (so that $Z_{H_{u,d}}\neq 0$) is insufficient to ensure a non-zero SD coupling.  If $M_{1}$, $M_{2} \rightarrow \infty$, so that a pure Higgsino is recovered,  $|Z_{H_u}| =|Z_{H_d}|=1/\sqrt{2}$, and the SD coupling vanishes.  Instead, the Higgsino forms a Dirac state, and the large vector scattering of the Dirac neutrino is recovered.  Hence, mixing with $\tilde{B}$ and/or $\tilde{W}$ (so that $|Z_{H_u}|\neq |Z_{H_d}|$) is required in order for the $d_q$'s to be non-zero.  This requirement also implies a non-zero SI cross section, giving the correlation demonstrated below.

The typical cross section for SD DD in the MSSM (again see Sec. \ref{sec:ArgueWellTempered}) is given by
\be\label{eq:SDinMSSM}
\sigma_\mr{SD}^\mr{MSSM}(\chi\,p\rightarrow \chi\,p) \approx 4 \times 10^{-4}\, \mr{pb}\, \left(\frac{|Z_{H_d}|^2-|Z_{H_u}|^2}{0.1}\right)^2 \qquad \textrm{(MSSM: SD typical)}.
\ee

There are reasons to expect the squarks do not make a sizable contribution to the DD cross sections.  In the MSSM, satisfying the LEP bound on the Higgs boson mass requires large radiative corrections from the stop loops.  This implies that at least one stop must have a TeV scale mass.  Renormalization group flow tends to make the third generation sparticles lighter than the partners for the first and second generations.  Therefore, it is plausible that squark contributions to DD scattering are negligible since only the first and second generation squarks contribute (see Appendix \ref{sec:SquarkExchange} for details about squark exchange).  For concreteness, in all scans below we take the scalar superparters to be $\mathcal{O}(2\,\mr{TeV})$.  This is also why Eqs.~(\ref{eq:SIinMSSM}) and (\ref{eq:SDinMSSM}) are expected to be good approximations.  For a study which focuses on the effects of light squarks, see \cite{Belanger:2008gy}.

\section{The Argument for a Well-Tempered Neutralino}\label{sec:ArgueWellTempered}
Arkani-Hamed, Delgado and Giudice \cite{ArkaniHamed:2006mb} argued that when one takes the LEP limits on charginos and sleptons into account, a pure neutralino (\emph{i.e.} composed of only one gaugino eigenstate, usually taken to be Bino) is no longer the ``natural" MSSM DM candidate, at least when one imposes the requirement of a thermal cosmology.  They claim that one should instead consider a mixed neutralino, which they have dubbed ``well-tempered."  Since the relic density of mixed DM is set by annihilations to $W^+\,W^-$ (and $t\,\bar{t}$ when kinematically allowed) there is a further condition that $m_{\chi} > m_W$.  Hence, we will impose this requirement when we refer to ``thermal" DM in the analysis that follows.  In what follows, we review their argument and then discuss some non-thermal options.  Note that SI DD has previously been studied for well-tempered models \cite{Baer:2006te, Hisano:2009xv}, but no dedicated SD study exists.

\subsection{Thermal history}
We begin by considering the thermal history of a nearly pure Bino.  If one does not allow for co- \cite{Griest:1990kh, Ellis:1998kh, Ellis:1999mm} or resonant \cite{Griest:1990kh, Drees:1992am, Roszkowski:2001sb, Djouadi:2005dz, Nath:1992ty} annihilations, then Bino freeze-out is controlled by $t$-channel sfermion exchange.  One can show \cite{ArkaniHamed:2006mb} that in order to produce the observed DM relic density, the sfermion must be  $\,\lsim 110$ GeV.  Since the LEP limits on sfermions are $\mathcal{O}(100\,\mr{GeV})$, there is only a small experimentally allowed window for thermal Bino DM.

Either co-annihilations (\emph{e.g.} with the stau or stop) or resonant annihilation through the pseudo-scalar Higgs ($A^0$) also allow dominantly Bino DM.  However, both of these options involve numerical coincidences.  In the first case the Boltzmann factor will exponentially suppress the density of the would-be co-annihilator unless $\mr{exp}(-\Delta M/T_{f})$ is $\mathcal{O}(1)$, where $\Delta M = m_\mr{NLSP} - m_{\chi}$, $m_\mr{NLSP}$ is the mass of the next-to-lightest superpartner, and $T_{f}$ is the DM freeze-out temperature.  Since $T_{f} \approx m_{\chi}/20$, this requires a mass degeneracy, $\Delta M$, of a few percent.  To realize the second case requires a precise relationship between $m_{\chi}$ and $m_{A}$.  When $m_{\chi} < m_W$, the $Z^0$ or $h$ poles may be used to achieve the correct relic density, which requires a similar numerical conspiracy.

Located at the other extreme, far away from the pure Bino, is a pure Wino or a pure Higgsino. In these cases, the requirement of a thermal relic abundance fixes the mass to be $\mathcal{O}(2.5\,\mr{TeV})$ and $\mathcal{O}(1\,\mr{TeV})$ respectively.  Thus,  to realize either of these cases implies $\mu \gtrsim \mathcal{O}(100\,\mr{GeV})$.  Since, in the MSSM, the $Z^0$ mass is given by 
\be\label{eq:mZinMSSM}
\frac{m_Z^2}{2} = - |\mu|^2 +\frac{m^2_{H_d}-m^2_{H_u} t^2_{\beta}}{t^2_{\beta} - 1},
\ee
where $m^2_{H_{u,d}}$ are the Higgs soft-mass squared parameters, this requires a substantial fine-tuning between $\mu^2$ and $m_{H_{u,d}}^2$ in order to reproduce the measured $Z^0$ mass of 91 GeV.  Therefore, the desire to alleviate fine-tuning in this expression leads to the requirement that $\mu \sim \mathcal{O}(100\,\mr{GeV})$.   This will also naively lead to well-tempering since the neutralino mixing is proportional to $m_Z/\mu$.  Though the accuracy of the current measurement of the DM relic density (see Eq.~(\ref{eq:OmegaDMWMAP})) requires a precisely determined neutralino composition, one can easily reproduce the DM abundance for any mass of ${\mathcal O}$(100 GeV).  The Bino/Higgsino mixed LSP as a good thermal WIMP was pointed out in studies of the focus point region on the MSSM \cite{Feng:2000zu, Feng:2000gh}.

A Higgs boson mass above the LEP bound requires large radiative corrections from a stop squark.  This implies that the scale for these particles, $m_\mr{SUSY}$, should be around a TeV.  These states yield additive corrections to $m^2_{H_{u,d}}$, proportional to $m^2_\mr{SUSY}$.  Hence, even in the case when $\mu \sim \mathcal{O}(100$ GeV), there will naively be fine-tuning between these corrections and the bare value of $m^2_{H_{u,d}}$ in order to reproduce $m_Z$.  Solutions to this ``little hierarchy problem" have been proposed within the MSSM (\emph{e.g.} \cite{Kitano:2005wc}) -- we will ignore this type of fine-tuning in our arguments, focusing instead on the model independent tuning explicit in Eq.~(\ref{eq:mZinMSSM}).

\subsection{Non-thermal options}
A thermal history is not the only way to achieve the correct DM relic abundance \cite{Moroi:1999zb}.  It has even been argued \cite{Acharya:2009zt} that there is a ``non-thermal WIMP miracle" when there exist TeV scale states which decay to the DM via Planck suppressed operators.  For example, a heavy gravitino (or string-theory moduli fields) can live long enough to dominate the energy density of the universe.  Then when these states decay, they will produce superpartners which will decay down to the lightest neutralino, resulting in a neutralino relic density.  This relaxes the relationship between the mass/composition and relic density of a neutralino.    

A variety of other options have been proposed.  Models where the energy density of the universe at the epoch of DM freeze-out was dominated by something other than radiation were studied in \cite{Kamionkowski:1990ni}.  Alternately, if the DM interacts so feebly that it never achieves thermal equilibrium, one can achieve the correct value of the relic density via ``freeze-in" production \cite{Hall:2009bx}.  Since the total energy density of DM is close to that of the baryons, one can construct models where the DM relic density is set by an asymmetry which is determined by the baryon asymmetry \cite{Kaplan:2009ag}.  In \cite{Gelmini:2006pw}, it was shown that by varying the reheat temperature and allowing for non-thermal sources, any neutralino composition can result in the correct relic density.  In \cite{Cohen:2008nb}, a low temperature phase transition in the early universe changes the DM properties after freeze-out.  All of these options involve either non-trivial cosmological histories or other model building challenges.  We will focus on the thermal -- and hence well-tempered -- case, with discussions of the deviations that arise when the thermal assumption is relaxed.  
 
\section{Spin Dependent Cross Sections for Mixed Dark Matter}\label{sec:SDphys}
In the MSSM, the neutralino mass mixing can often be approximately understood in terms of a two state system:  a Dirac Higgsino mixing with either a Bino or a Wino.   Thus, to understand the physics of SD scattering via $Z^0$ exchange, it is useful to consider the simple ``Singlet-Doublet Model" (SDM) for DM, where the singlet has the same quantum numbers as either a Bino or a Wino, and the doublets have the same quantum numbers as the Higgsinos:
\begin{equation}\label{eq:L_SDM}
{\mathcal L_\mr{SDM}} \ni \mu_{D}\, D\, \bar{D} + \lambda\, \textit{\textbf{h}}\, S\, D + \lambda^{\prime}\, \textit{\textbf{h}}^{\ast}\, S\, \bar{D} + \frac{\mu_S}{2}\, S^{2}.
\end{equation}
Here $D$ and $\bar{D}$ are a vector-like pair of $SU(2)$ doublet fermions, $S$ is an $SU(2)$ singlet, $\textit{\textbf{h}}$ is the SM Higgs doublet, $\lambda$ ($\lambda'$) is the Yukawa coupling which leads to the mixing between the $D\, (\bar{D})$ and $S$, $\mu_D$ is the vector-like mass for the $D$ and $\bar{D}$, and $\mu_S$ is the Majorana mass for $S$.  For the purposes of SD scattering it is sufficient to replace $\textit{\textbf{h}}$ by its vev, $\langle \textit{\textbf{h}} \rangle \equiv v = 174$ GeV.  The exchange of the uneaten component of $\textit{\textbf{h}}$ leads to SI DD.    

In the case where $S$ plays the role of the Bino, the values of $\lambda$ and $\lambda^{\prime}$ are constrained by the supersymmetric relations to be $\lambda\,v=-m_Z \,s_w\,c_{\beta}$ and $\lambda'\, v=-m_Z\,s_w\,s_{\beta}$, while in the case where $S$ is the Wino, the values of $\lambda$ and $\lambda^{\prime}$ are constrained by the supersymmetric relations to be $\lambda\, v=m_Z\,c_w\,c_{\beta}$ and $\lambda'\, v=m_Z\,c_w\,s_{\beta}$.   

We now use this model to discuss the coupling of the $Z^0$ boson to the DM in the MSSM.  In Appendix \ref{sec:BinoAndWinoLimit} we discuss the diagonalization of the $3 \times 3$ mixing matrix of the SDM.  With appropriate substitutions, these expressions correspond to either Bino/Higgsino ($M_2\rightarrow \infty)$ and Wino/Higgsino $(M_1\rightarrow \infty)$ neutralinos.  In these limits we can write down approximate expressions for the effective coupling of the DM to the $Z^0$.  When there are no degeneracies between parameters in the neutralino mass matrix and $m_Z$ may be treated as a perturbation, we have (see \cite{ArkaniHamed:2006mb} and Appendix \ref{sec:BinoAndWinoLimit}):
\bea |Z_{H_d}|^2-|Z_{H_u}|^2 = \left\{ \begin{array}{ll}
\frac{c_{2\beta}\,s_w^2\,m_Z^2}{\mu^2-M_1^2} & \mbox{for $|M_1|,\,|\mu|,\,|\mu|-|M_1|>m_Z$, $M_2\rightarrow \infty$}\\
\frac{c_{2\beta}\,c_w^2\,m_Z^2}{\mu^2-M_2^2} & \mbox{for $|M_2|,\,|\mu|,\,|\mu|-|M_2|>m_Z$, $M_1\rightarrow \infty$}.\end{array} \right. 
\eea
The largest values of $ |Z_{H_d}|^2-|Z_{H_u}|^2$ do not occur in this limit.  Instead, they are found when two parameters of the neutralino mass matrix are degenerate.  The reason is simple:  a degeneracy allows a large gaugino--Higgsino mixing in spite of the relative smallness of the off-diagonal entries of the neutralino mass matrix (proportional to $m_{Z}$).  It should be said that there is no particular reason to believe that a precise degeneracy should occur, since $\mu$ and the gaugino masses are SUSY preserving and breaking respectively.  However, since this case maximizes the possible signal at SD experiments, it is worth noting.  In the presence of these degeneracies, we have (see Appendix \ref{sec:BinoAndWinoLimit}):
\bea |Z_{H_d}|^2-|Z_{H_u}|^2 = \left\{ \begin{array}{ll}
\frac{(s_{\beta}-c_{\beta})\,s_w\,m_Z}{2\sqrt{2}\,|\mu|}+\frac{(s_{\beta}^2-c_{\beta}^2)\,s_w^2\,m_Z^2}{8\,\mu^2} & \mbox{for $|M_1| = |\mu| > m_Z$, $M_2\rightarrow \infty$} \\
\frac{(s_{\beta}-c_{\beta})\,c_w\,m_Z}{2\sqrt{2}\,|\mu|}+\frac{(s_{\beta}^2-c_{\beta}^2)\,c_w^2\,m_Z^2}{8\,\mu^2} & \mbox{for $|M_2| = |\mu| > m_Z$, $M_1\rightarrow \infty$} \label{eqn:muM2degen}.\end{array} \right.  \eea 
Perturbing away from the limit of exact degeneracy gives corrections to these expressions of ${\mathcal  O} ((M_{i} - \mu)/\mu)$.  Note that DM with a mixed Wino/Higgsino  has a SD DD rate enhanced relative to a Bino/Higgsino admixture by the appropriate power of $c_w/s_w = 1.8$.

What is the largest obtainable SD cross section in the MSSM?  A numerical scan yields  
\bea
|Z_{H_d}|^2-|Z_{H_u}|^2 &<&0.4 \Rightarrow\\
(\sigma_\mr{SD}^\mr{SUSY})  &<& 6\times 10^{-3}\,\mr{pb} \qquad \rm{(General\; MSSM, \; Non-thermal \;DM)},\label{eq:sigmaSDMaxSUSY}
\eea
when the squarks are heavy.   This upper bound is largely a consequence of   the LEP bounds on the chargino masses which force the mixing $\sim m_Z/\mu$ to be less than one.  Eq.~(\ref{eqn:muM2degen}) provides a good analytic understanding of this number -- it comes within approximately 10\% of this value.  The deviation is due to mixing effects that occur away from the large $M_{1}$ limit. 

In many models of SUSY breaking the relation $M_1/\alpha_1=M_2/\alpha_2=M_3/\alpha_3$  holds.  We refer to this condition as unified gaugino masses.  Because this is equivalent to $M_2 \approx 2 M_1$ at the weak scale, the LSP is mostly Bino and Higgsino.  In this case, 
\bea
|Z_{H_d}|^2-|Z_{H_u}|^2 &<& 0.32 \Rightarrow \\
(\sigma_\mr{SD}^\mr{SUSY})  &<&  4 \times 10^{-3}\,\mr{pb} \qquad \rm{(Unified \; Gaugino\; Masses, \; Non-thermal\; DM)}.\label{eq:SDMaxUnifiedGauginos}
\eea

Finally, for $m_\chi > m_W$, a thermal relic density within $\pm 3\,\sigma$ of the WMAP measurement implies an upper limit on the amount of Higgsino in the DM particle.  Therefore,
\bea
|Z_{H_d}|^2-|Z_{H_u}|^2 &<& 0.24 \Rightarrow  \\
(\sigma_\mr{SD}^\mr{SUSY})_\mr{thermal}  &<& 2 \times 10^{-3}  \, \rm{pb} \qquad \rm(General\; MSSM, \;Thermal\; DM).\label{SDMaxThermal} 
\eea
This result holds for the case with unified gaugino masses as well.  Note that Eqs.~(\ref{eq:sigmaSDMaxSUSY}), (\ref{eq:SDMaxUnifiedGauginos}), and (\ref{SDMaxThermal}) all occur for a DM mass of $\mathcal{O}(80\,\mr{GeV})$.

To saturate the above bound (\emph{i.e.} maximize $\sigma_\mr{SD}$ for thermal, well-tempered DM) requires a Bino/Higgsino mixture (recall that $d_{q}$ vanishes for a pure Higgsino), with a negligible Wino contribution.  The largest values of SD DD occur when the DM has the largest Bino/Higgsino mixing which happens for the lowest values of the DM mass.  As the mass of the DM increases, a larger component of Higgsino or Wino is needed for the DM to efficiently annihilate down to the correct relic density, which in turn typically leads to a decrease in $\sigma_\mr{SD}$.

\begin{figure}[h!]
\begin{center}
\includegraphics[width=1.0\textwidth]{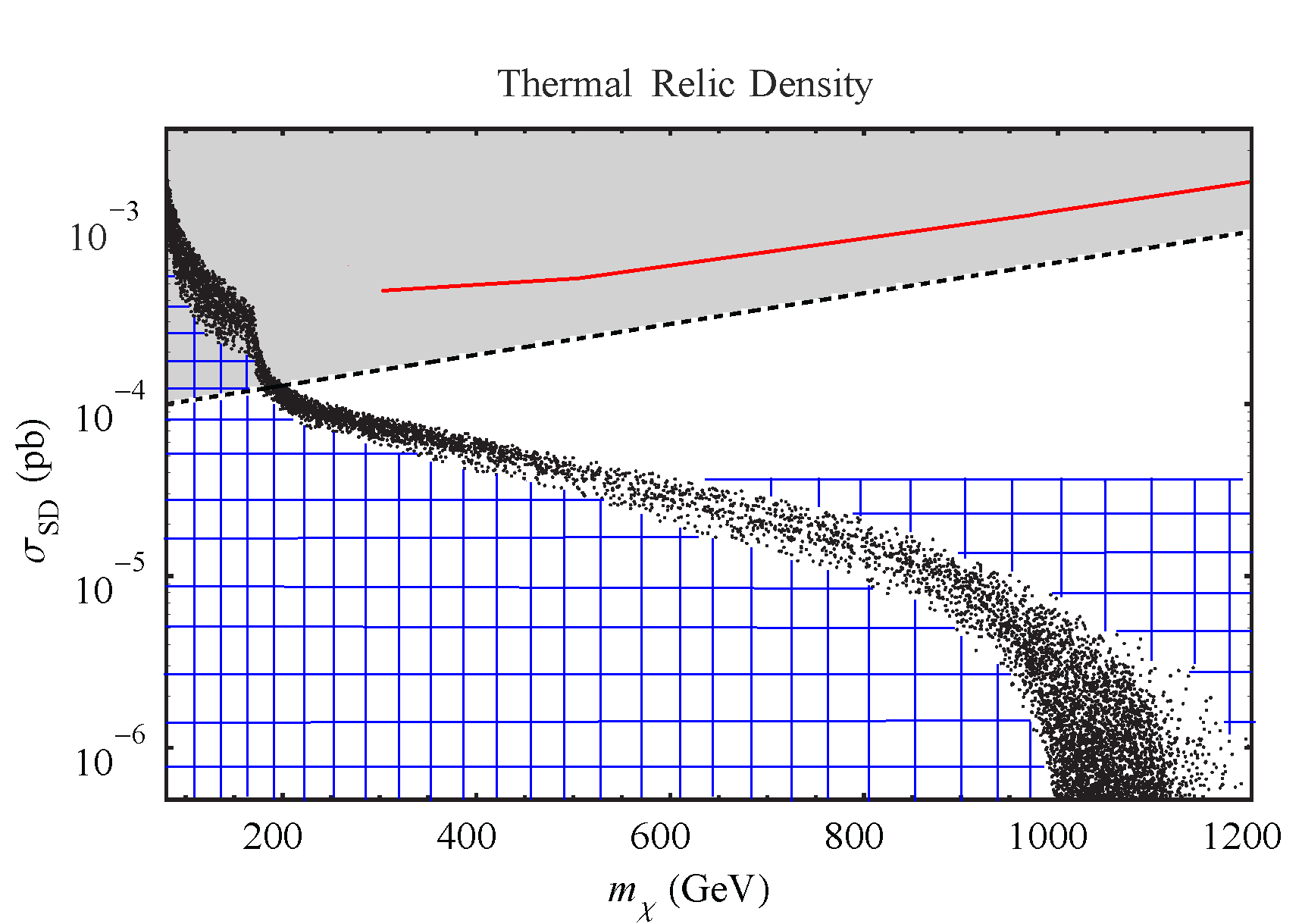} 
\end{center}
\caption{$\sigma_\mr{SD}^p$, as a function of $m_\chi$ for points satisfying the relic density constraint.  We have imposed gaugino mass unification and taken the decoupling limit.  The shaded region above the dotted line corresponds to ``large" SD and will be probed in the near term.  The solid red line is the current bound from IceCube, assuming annihilation to $W^+\,W^-$.  The blue hatched region is filled in if the assumption of gaugino mass unification is relaxed.  The sfermion masses are taken to be $\mathcal{O}(2\,\mr{TeV})$.}
\label{fig:maxSDforRelicdensityUnifiedGauginos}
\end{figure}

As shown in Fig.~\ref{fig:maxSDforRelicdensityUnifiedGauginos}, there is a tight correlation between the SD cross section and the DM mass, in the decoupling limit when there is gaugino mass unification and a thermal relic abundance.

For low masses, the neutralino is well-tempered for low masses and as $m_{\chi}\rightarrow \mathcal{O}(1\,\mr{TeV})$ the neutralino approaches a pure Higgsino. Examining  Fig.~\ref{fig:maxSDforRelicdensityUnifiedGauginos}, except for when the annihilation channel $\chi\chi \to t\bar{t}$ opens, $\sigma_\mr{SD}$ is a smooth, monotonically decreasing curve.  An experiment sensitive to cross sections of $\mathcal{O}(10^{-4}\,\mr{pb})$ will probe  $m_\chi \lesssim 200$ GeV.  There is a spread in the points in this figure from the liberal range taken on the relic density constraint.  For masses approaching $\mathcal{O}(1\,\mr{TeV})$, there is additional extent from the variation in the Bino content of the neutralino and from contributions from squark exchange.  For masses at 1 TeV, $\sigma_\mr{SD}$ goes from $10^{-6}\,\mr{pb}\,\rightarrow 0$ for $M_1$ from $1300\,\mr{GeV}\rightarrow \infty$.  Note that the projected reach of a 1 ton version of COUPP is $\mathcal{O}(10^{-6})$ pb for $m_{\chi} = 1$ TeV \cite{Bertone:2007xj}, \emph{which would probe the entire range of $\mr{SD}$ cross sections for neutralinos excepting a nearly pure $\mr{TeV}$ Higgsino}. 

Note that the imposition of the unified gaugino mass condition essentially imposes the requirement that there is a tiny Wino content in the LSP.  The hatched region in Fig.~\ref{fig:maxSDforRelicdensityUnifiedGauginos} is filled in when non-unified gaugino masses are allowed.  In this case, a thermal relic DM candidate can be obtained for a Bino tempered with Wino if $M_1 \approx M_2$, which implies that the SD cross section decreases, effectively filling in the region beneath the curve in Fig.~\ref{fig:maxSDforRelicdensityUnifiedGauginos}.  Note that when $\sigma_\mr{SD}\sim \mathcal{O}(10^{-6}\,\mr{pb})$, there is additional model dependence since the squark contribution becomes important (see Appendix \ref{sec:SquarkExchange}).  

Finally, we note that there is a region of well-mixed Higgsino--Wino near 2 TeV with a thermal abundance (where $M_{2} \approx \mu$).  In this case, the second line of Eq.~(\ref{eqn:muM2degen}) applies, and we find an approximate SD cross section of $6 \times 10^{-6}$ pb, perhaps able to be probed at a future 1 ton COUPP-like experiment.  These are the neutralinos which account for the hatched region above the points in Fig.~\ref{fig:maxSDforRelicdensityUnifiedGauginos}.

Not only is the SDM a simplified system useful for understanding the physics of SD scattering in the MSSM, it is potentially of independent interest.  The DM may be unrelated to the solution to the hierarchy and simply given by the Lagrangian of  Eq.~(\ref{eq:L_SDM})  \cite{ArkaniHamedFN, SenatoreMahbubani}.  Then the DD story is essentially unchanged except there is greater parametric freedom.

For example, the Higgs boson mass is no longer fixed by SUSY.  Then the only constraint is $ m_h \lesssim\mathcal{O}(\mr{TeV})$ to unitarize $W^{\pm}_L$ scattering.  For $m_h\sim$ TeV, the SI DD cross section is at most $10^{-12}\,\mr{pb}$ which would not lead to a signal in the next round of SI experiments.   While such a large Higgs boson mass is in tension with precision electroweak measurements, it could be reconciled with a contribution to the $T$ parameter \cite{PeskinWells} in a way that factorizes from the DM phenomenology.  

If one allows for a non-thermal history, the freedom of the SDM allows off-diagonal parameters of the mixing matrix that give $|Z_{H_d}|^2-|Z_{H_u}|^2 = 1$.  This maximizes the SD DD signal from $Z^0$ exchange ($\sigma_\mr{SD}^\mr{SDM}  \approx 4\times 10^{-2}$ pb).  Thus, the SDM with a non-thermal history predicts scattering anywhere up to (or even above) the current bounds.  Requiring a thermal history limits the amount of doublet allowed in $m_\chi$, decreasing $\sigma_\mr{SD}^\mr{SDM}$.  For if a very large doublet component is chosen (in an attempt to maximize the SD cross section), the requirement of reproducing the relic density requires $\mu_S$ to be $\mathcal{O}(\mr{TeV})$.

\section{Spin Independent versus Spin Dependent}\label{sec:SIvsSD}
When a Majorana fermion couples to the $Z^0$, there is necessarily an interaction with a Higgs boson, which leads to SD and SI elastic scattering respectively.  In the last section, we concentrated on the physics behind the size of the SD cross section.  We now ask the following questions:  what is the expected correlation between the SI and SD signals?   Is it possible to make one large while the other nearly vanishes?

Since $m_h$ and $m_Z$ are known in the MSSM, there exists a correlation between the SI and SD signals, at least in the limit of heavy sfermions and Higgs boson decoupling.  For this region of MSSM parameter space, the SI and SD DD cross sections are given by Eqs.~(\ref{eq:SIinMSSM}) and (\ref{eq:SDinMSSM}), where only mixing factors and the Higgs boson mass are left unspecified.  The light Higgs boson mass is constrained to lie in the tight range 114 GeV $< m_h <$ 130 GeV, where the lower bound is due to the LEP limit and the upper bound comes from considerations of fine-tuning.  For the SplitSUSY model -- where the decoupling and heavy sfermion limits certainly apply -- the Higgs boson mass is allowed to be larger: $m_h < 160$ GeV.

\begin{figure}[h]
\begin{center}
\includegraphics[width=1.0\textwidth]{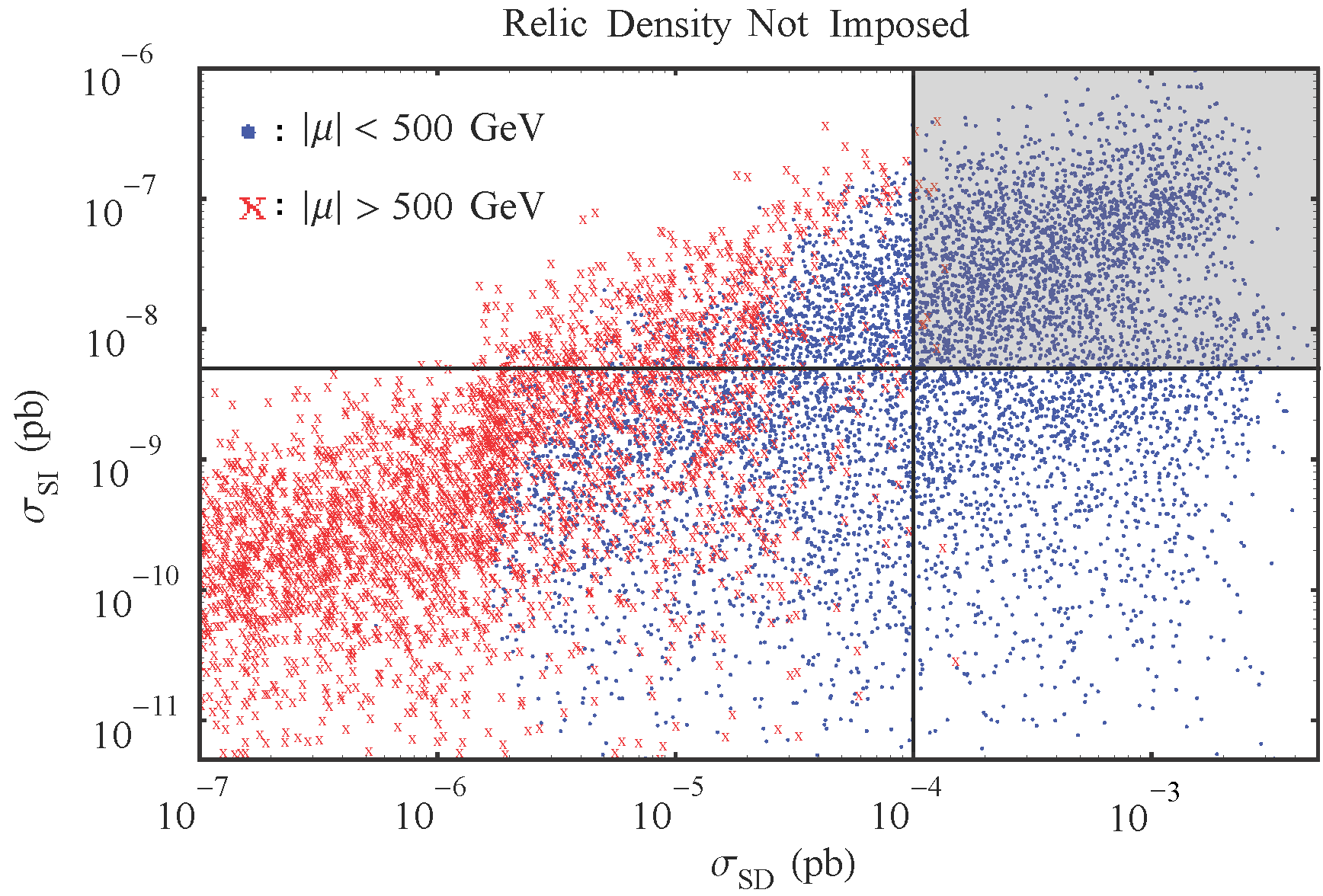}
\end{center}
\caption{The max$(\sigma_\mr{SI}^p,\,\sigma_\mr{SI}^n)$ vs. $\sigma_\mr{SD}^p$ cross sections in pb for the MSSM.  The dots (in blue) and crosses (in red) correspond to $|\mu| < 500$ GeV and $|\mu| > 500$ GeV respectively.  The horizontal (vertical) line refers to the projected sensitivity for the next generation of SI (SD) experiments. We have shaded the near-term probeable region.  Note that we are neglecting the dependence of this sensitivity on the neutralino mass.  We have \emph{not} imposed the thermal relic density constraint -- all points are taken to have $\rho_\mr{DM}=0.3 \,\mr{GeV/cm}^{3}$, regardless of thermal abundance.  All sfermions have masses of $\mathcal{O}$(2 TeV).  If one takes the decoupling limit, there is a maximum value for $\sigma_\mr{SD} = 3 \times 10^{-8}$ pb.}
\label{fig:SIvsSD}
\end{figure}

\begin{figure}[h]
\begin{center}
\includegraphics[width=1.0\textwidth]{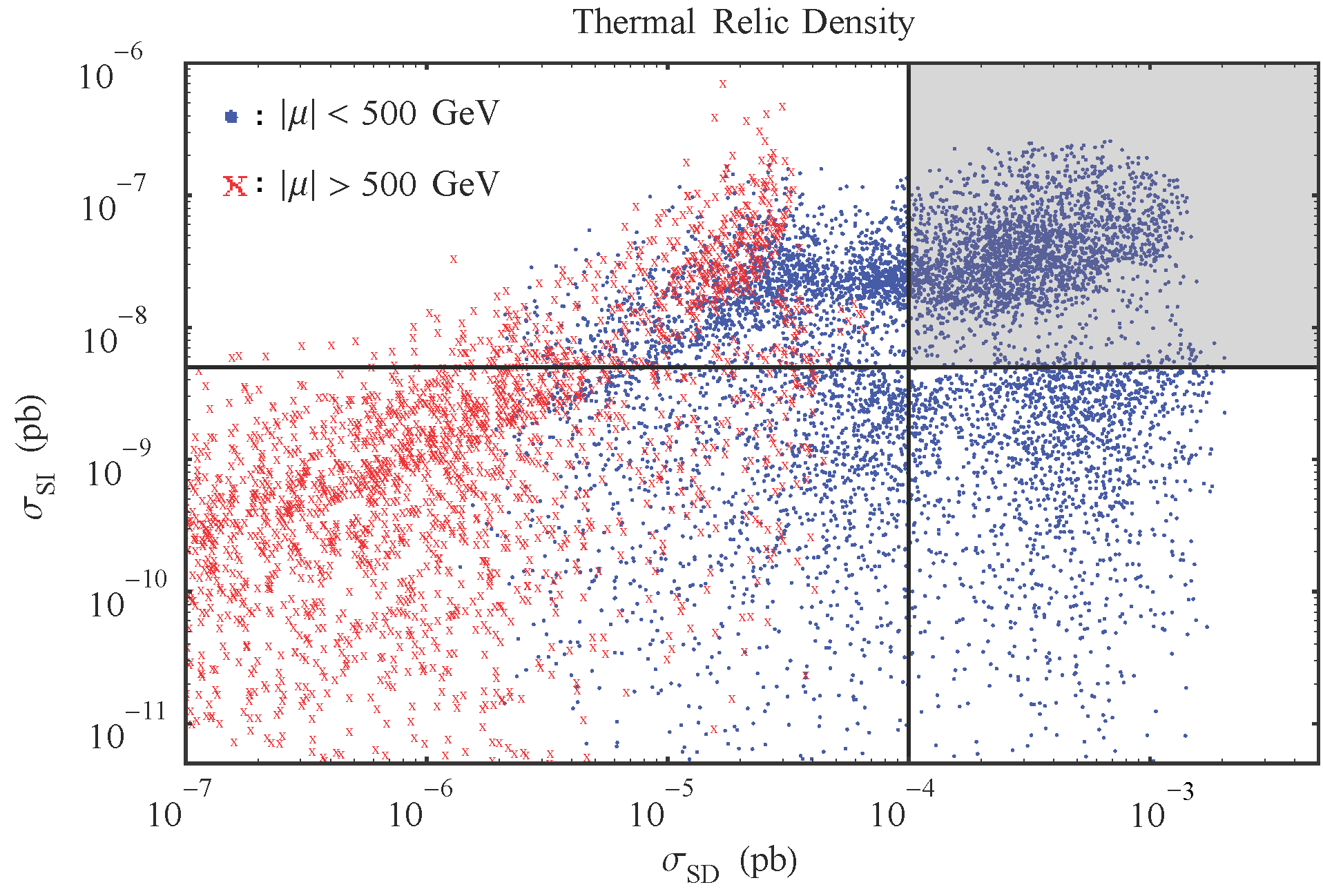}
\end{center}
\caption{The max$(\sigma_\mr{SI}^p,\,\sigma_\mr{SI}^n)$ vs. $\sigma_\mr{SD}^p$ cross sections in pb for the MSSM.  We have imposed that the thermal abundance of the neutralinos is within $\pm\,3\,\sigma$ of the WMAP measurement.  The dots (in blue) and crosses (in red) correspond to $|\mu| < 500$ GeV and $|\mu| > 500$ GeV respectively.  The horizontal (vertical) line refers to the projected sensitivity for the next generation of SI (SD) experiments.  We have shaded the near-term probeable region.  Note that we are neglecting the dependence of this sensitivity on the neutralino mass.  All sfermions have masses of $\mathcal{O}$(2 TeV).}
\label{fig:SIvsSDThermal}
\end{figure}

\begin{figure}[h]
\begin{center}
\includegraphics[width=1.0\textwidth]{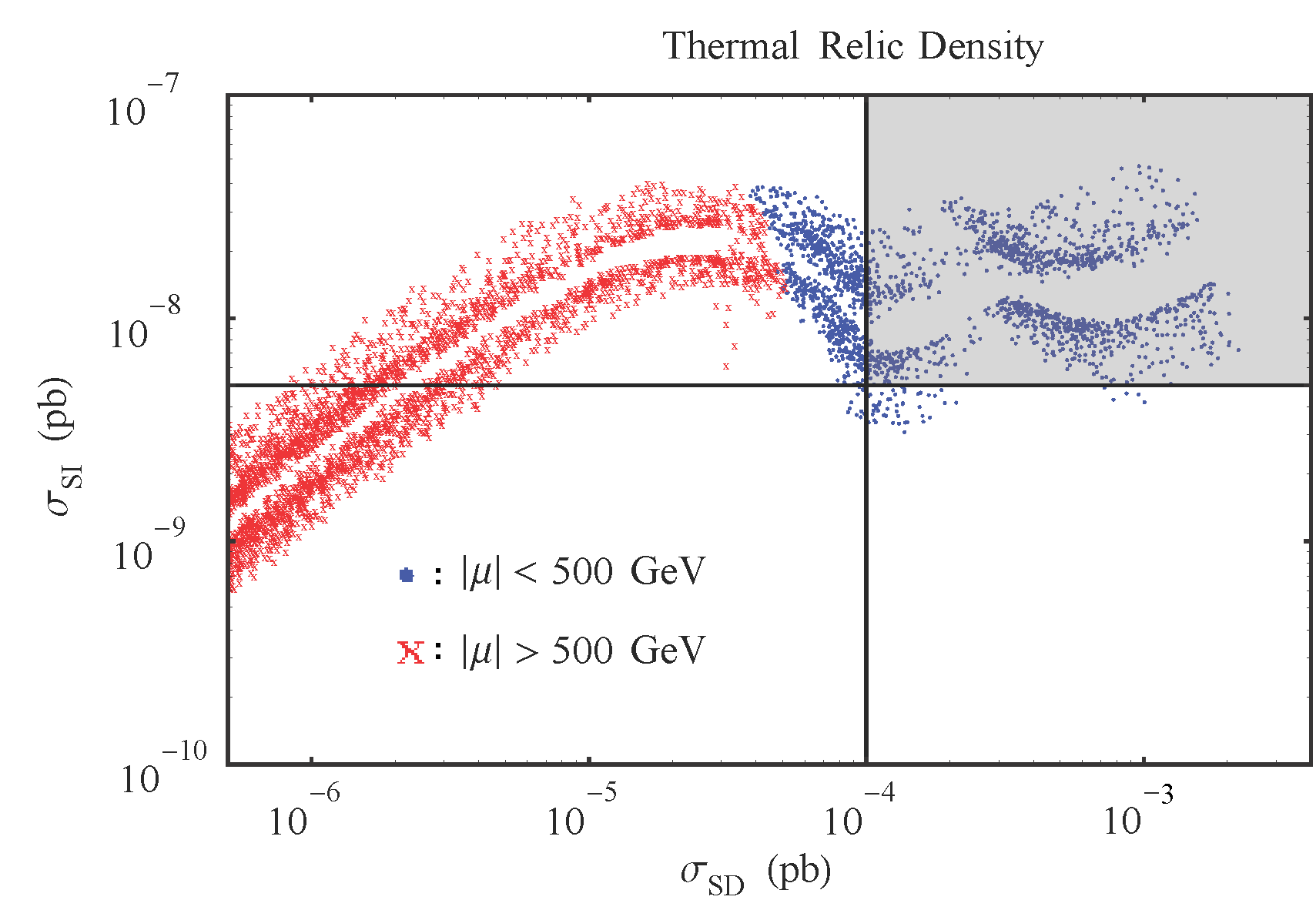}
\end{center}
\caption{The max$(\sigma_\mr{SI}^p,\,\sigma_\mr{SI}^n)$ vs. $\sigma_\mr{SD}^p$ cross sections in pb for the MSSM with gaugino mass unification.  We have imposed that the thermal abundance of the neutralinos is within $\pm\,3\,\sigma$ of the WMAP measurement.  We have taken the decoupling limit ($m_A = 4\,\mr{TeV}$).  The dots (in blue) and crosses (in red) correspond to $|\mu| < 500$ GeV and $|\mu| > 500$ GeV respectively (see the text for a discussion).  The horizontal (vertical) line refers to the projected sensitivity for the next generation of SI (SD) experiments.  We have shaded the near-term probeable region.  Note that we are neglecting the dependence of this sensitivity on the neutralino mass.  All sfermions have masses of $\mathcal{O}$(2 TeV).}
\label{fig:SIvsSDUnifiedGauginos}
\end{figure}

In Figs. \ref{fig:SIvsSD}, \ref{fig:SIvsSDThermal} and \ref{fig:SIvsSDUnifiedGauginos}, we have plotted the max$(\sigma_\mr{SI}^p,\,\sigma_\mr{SI}^n)$ vs. $\sigma_\mr{SD}^p$ for neutralino scattering with various restrictions.  Note that these plots are made from independent scans and we have taken the scalar superpartners to be $\mathcal{O}$(2 TeV).  

As discussed in Section \ref{sec:DD}, we define ``large" cross sections to be $\sigma_\mr{SI}^\mr{large} > 5\times 10^{-9}$ pb and $\sigma_\mr{SD}^\mr{large} > 10^{-4}$ pb, motivated by the projected near term range of current DD experiments.  Hence, the shaded region delineates the (very approximate) reach of the next generation of SI and SD experiments.  Note that this neglects the dependence of the sensitivity on the mass of the DM.  The maximum for $\sigma_\mr{SD}$ in Fig. \ref{fig:SIvsSD} is given by Eq.~(\ref{eq:sigmaSDMaxSUSY}) and for Figs.~\ref{fig:SIvsSDThermal} and \ref{fig:SIvsSDUnifiedGauginos} is given by Eq.~(\ref{SDMaxThermal}).  

In Fig. \ref{fig:SIvsSD} we show points for both thermal and non-thermal neutralinos.  This is our most general framework, and in this case it is clear that the correlation between the relevant mixing angles (and hence cross sections) is weak.  By only allowing points which have a thermal relic density within $\pm\,3\,\sigma$ of the WMAP measurement (see Figs. \ref{fig:SIvsSDThermal} and \ref{fig:SIvsSDUnifiedGauginos}), the correlation progressively improves.  We will discuss this in detail in what follows. 

We will pay special attention to the $m_{H} \rightarrow \infty$ limit. In any theory with multiple Higgs bosons, a small SI signal can occur when the diagrams from Higgs boson exchange cancel against one another.  Two important points should be made.  First, this cancellation is often incomplete and typically cannot be realized for scattering off of both protons and neutrons simultaneously.  Second, such a cancellation is a conspiracy -- it requires unexpected relationships between parameters in the Higgs sector and nuclear matrix elements.    The finer the cancellation, the greater the conspiracy  (for further discussion of this cancellation, see Sec.~\ref{sec:LargeSmall}).  If one takes the decoupling limit for Fig. \ref{fig:SIvsSD}, so that SI DD is determined by $h$ exchange alone, the maximum SI cross section is $\sim 3 \times10^{-8}$ pb.  Note that even for $m_A \sim \mathcal{O}(\mr{TeV})$ there can be nontrivial contributions for $t_{\beta} \sim \mathcal{O}(50)$ (see Eq.~(\ref{eq:cdDecoupling})). 

There is a negative correlation between fine-tuning and the size of DD cross sections  (see Eq.~(\ref{eq:mZinMSSM})) \cite{Mandic:2000jz,Kitano:2005ew}.   To emphasize this point, in Figs. \ref{fig:SIvsSD}, \ref{fig:SIvsSDThermal} and \ref{fig:SIvsSDUnifiedGauginos} we have marked points with $|\mu|<500$ GeV by blue dots and points with $|\mu|>500$ GeV by red crosses.   The apparent feature around $\sigma_\mr{SI} \approx 10^{-8}$ pb in Fig. \ref{fig:SIvsSDThermal} is due to the finite range of $m_A$ taken in this scan ($m_A < 1\,\mr{TeV}$) -- the points above this gap have constructive contributions from $h$ and $H$ while the points below have destructive contributions.  There are a few interesting features in Fig. \ref{fig:SIvsSDUnifiedGauginos}.  The gap which extends along the entire plotted range of SD cross sections is due to a slight cancellation between the various contributions from the light Higgs boson (see Eqs.~(\ref{eq:cuMSSM}) and (\ref{eq:cdMSSM})) which can occur at finite $t_{\beta}$ ($t_{\beta} < 50$ in this scan).  The small number of points around $\sigma_\mr{SD} = 3\times 10^{-4}\,\mr{pb}$ is due to the opening of the top threshold (see Fig.~\ref{fig:maxSDforRelicdensityUnifiedGauginos}).  The behavior around $\sigma_\mr{SD} = 2\times 10^{-5}\,\mr{pb}$ is due to the cross over from dominantly Bino to dominantly Higgsino DM, which occurs around $m_{\chi} = 500\, \mr{GeV}$.  

\subsection{Large SI and Large SD}

To have non-zero SI and SD signals, a Bino-Higgsino, Wino-Higgsino or Bino-Wino-Higgsino mix is required.  In fact, appreciable SI and large SD signals can be generated as long as the Higgsino fraction is larger than $\mathcal{O}(10\%)$.   Note that the $|\mu|<500\,\mr{GeV}$ points, which correspond to less fine-tuning in $m_Z$, imply large SD signals.  When the gaugino fraction is dominated by Wino rather than Bino, the relative size of $g$ and $g'$ gives a slight enhancement in the SI cross section.  There can be further enhancement of the SI cross section if $\mr{sgn}(Z_B)\ne\mr{sgn}(Z_W)$ (see Eq.~(\ref{eq:cuMSSM})) which accounts for points with the largest SI values in Figs. \ref{fig:SIvsSD} and \ref{fig:SIvsSDThermal}.  This cannot occur in models with unified gaugino masses, where $M_2 \approx 2\,M_1$.

Large SI and SD signals occur as long as there is non-trivial gaugino content in the WIMP.  Imposition of the thermal relic density constraint for $m_{\chi} > m_W $, ensures a minimum required Bino component.  If one imposes the large SI and SD conditions, $|Z_B|^2 \lesssim 0.7$ and $|Z_B|^2 \lesssim 0.85$ below and above the top threshold respectively.  Note that the large SD requirement implies that $m_{\chi} < 200\,\mr{GeV}$ (see Fig.~\ref{fig:maxSDforRelicdensityUnifiedGauginos}).  Hence, the assumption of a thermal history is necessary to conclude that the neutralino is a Bino-Higgsino admixture, rather than Wino-Higgsino.

In the next three subsections we will attempt to elucidate the difficulties one encounters when trying to suppress SI and/or SD.  This will allow us to argue that large SI and SD DD signals are the generic prediction for a well-tempered MSSM neutralino, since suppression of either SI or SD or both requires doing some gymnastics.  While future data may force  these contortions upon us, we conjecture that if the DM is a well-tempered neutralino, it is likely to be discovered in the next generation of DD experiments.

\subsection{Small SI and Small SD}
There are two ways to suppress both SI and SD.  The first is to make $|Z_{H_u}| = |Z_{H_d}| = 0$, which is equivalent to the $\mu\rightarrow \infty$ limit.  This limit leads to fine-tuning of the electroweak scale.  To achieve the proper thermal relic abundance in this case requires a  Bino-Wino mix.  Note that the Bino and Wino only mix indirectly through the Higgsino.  Therefore, two insertions of the mixing factor are required, and the resulting mixing is of size $(m_Z/\mu)^2$.
 One can see the effects of this limit by inspecting the red crosses in Figs.~\ref{fig:SIvsSD}, \ref{fig:SIvsSDThermal} and \ref{fig:SIvsSDUnifiedGauginos}.  The upper bound in Figs.~\ref{fig:SIvsSD} and \ref{fig:SIvsSDThermal} are from points which are either Bino/Higgsino or fully mixed states while the points with the smallest values for SI are due to either Wino/Bino neutralinos or the cancellations discussed in Sec.~\ref{sec:SmallSILargeSD}.  

The second option is to take $M_{1,2}  \gg \mu$.   This will imply that $Z_{B,W} = 0$, thereby suppressing SI DD, and $|Z_{H_u}| = |Z_{H_d}|$ so that SD DD is also zero.  Reproducing the measured relic density then requires $\mu \approx$ 1 TeV.  When one does impose the thermal relic density as a prior, Fig.~\ref{fig:maxSDforRelicdensityUnifiedGauginos} shows that for DM masses of $\mathcal{O}(\mr{TeV})$, \emph{i.e.} the region of dominantly Higgsino DM, the SD cross section ranges from $\mathcal{O}(10^{-5}\,\mr{pb})$ to 0.  Fig.~\ref{fig:SIvsSDUnifiedGauginos} shows the corresponding SI cross sections for this range.  The trend of SI and SD going to zero in this plot is due to the limit $M_{1,2}\rightarrow \infty$.  Thermal dark matter in either of these two limits ($\mu$ or $M_{1,2} \rightarrow \infty$) will have a finely-tuned electroweak scale.  Note that for either pure Wino or pure Higgsino DM there is a 1-loop diagram which leads to an SI DD cross section of $\mathcal{O}(10^{-11}\,\mr{pb})$ or $\mathcal{O}(10^{-12}\,\mr{pb})$ and an SD DD cross section of $\mathcal{O}(10^{-9}\,\mr{pb})$ or $\mathcal{O}(10^{-10}\,\mr{pb})$ for the Wino or Higgsino case respectively \cite{Hisano:2004pv}.  We neglect this tiny contribution in our numerical scans.

\subsection{Large SI and Small SD}\label{sec:LargeSmall}
There are points which have large SI and SD with a nearly maximal gaugino fraction.  If one relaxes the requirement of large SD, then the gaugino fraction can be pushed to nearly 100\% while keeping the product $Z_{B,W}\,Z_{H_{u,d}}$ approximately fixed, which in turn keeps the SI cross section constant.  The relic density constraint can still be satisfied since both Winos and Higgsinos annihilate to $W^{\pm}$ bosons with approximately the same rate.

There is another way to have small SD while allowing large SI.  In the context of the SDM, one can take $\lambda = \lambda'$, \emph{i.e.} $t_{\beta} = 1$ in the MSSM.  From the SDM mass matrix (see Appendix~\ref{sec:BinoAndWinoLimit}), one can see that mixing between $S$ and $D_-$ will vanish.  Since the SD cross section is proportional to this mixing factor, $Z_{D_-}$, it will be zero as well.  This effect accounts for the empty region in Figs.~\ref{fig:SIvsSD} and \ref{fig:SIvsSDThermal} since we restricted $t_{\beta} > 5$ in our numerical scans.

For $\tan \beta \gsim 1.5$, we find that for $\sigma_\mr{SI}\sim 5\times 10^{-9}$ pb the smallest cross section for SD is $\sigma_\mr{SD}\sim 10^{-6}$ pb.  If one allows $\sigma_\mr{SI} < 5\times 10^{-9}$ pb, then as $|Z_{H_{u,d}}| \rightarrow 0$, $\sigma_\mr{SD}/\sigma_\mr{SI} \rightarrow |Z_{H_{u,d}}|^2 \rightarrow 0$.  Hence, SD falls off faster than SI.  However, this is the $\mu\rightarrow \infty$ limit which leads to fine-tuning as described above.

\subsection{Small SI and Large SD}\label{sec:SmallSILargeSD}
Large SD requires a well-tempered neutralino, which naively also leads to large SI DD.  In this section we will enumerate the various options one has for suppressing SI signals.  We will argue that all options require fine-tuning or numerical coincidences\footnote{Another possibility is that both SI and SD from exchange of the $Z^0$ and Higgs boson respectively are small.  If there exist light squarks, they can give rise to large SD signals \cite{Bertone:2007xj}.  Cross section estimates from light squark exchange are discussed more in Appendix \ref{sec:SquarkExchange}.}.

\begin{figure}[h]
\begin{center}
\includegraphics[width=1.0\textwidth]{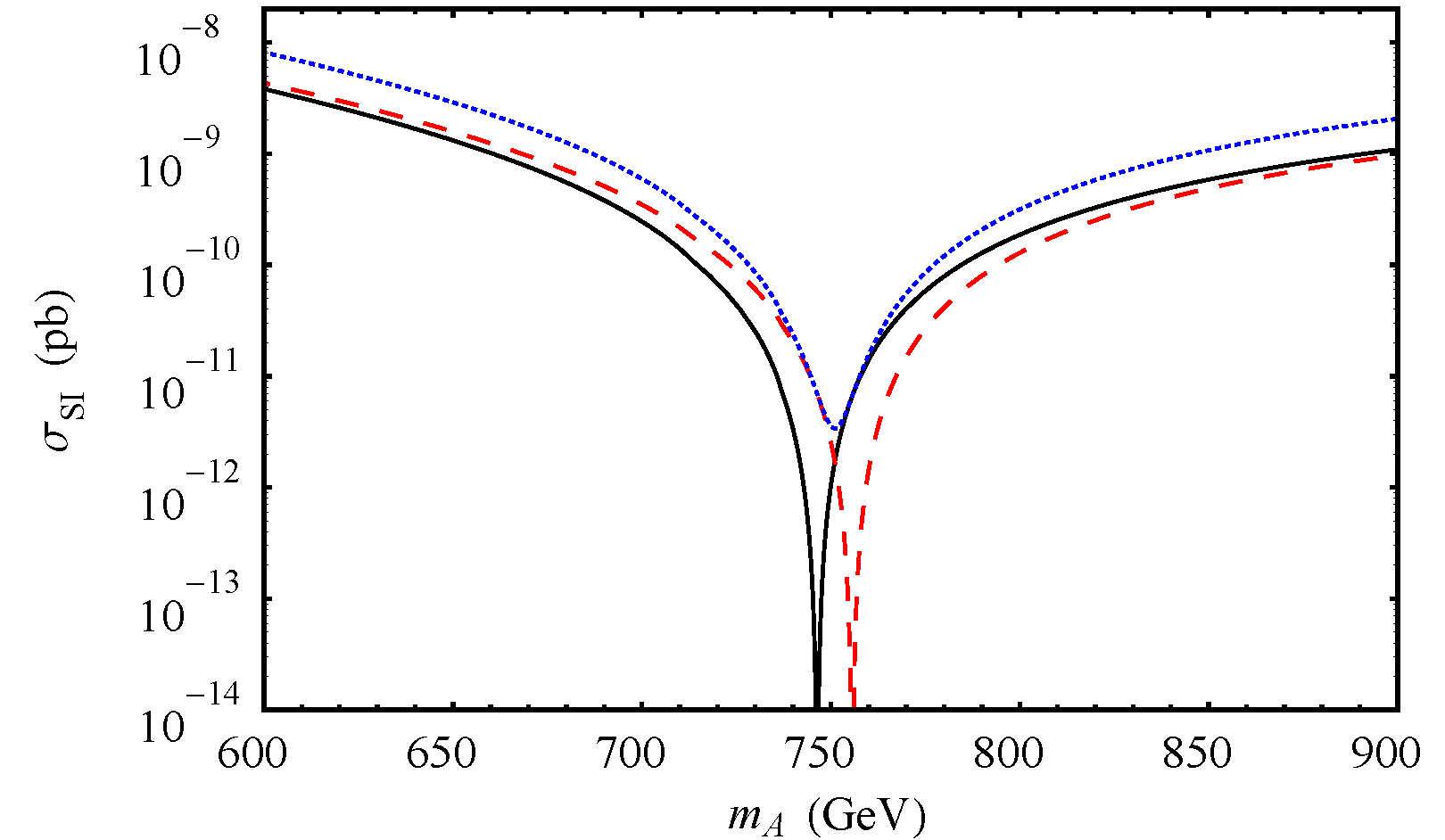}
\caption{Plot of the SI DD cross section for the neutralino scattering off of a proton (solid), a neutron (dashed) and both (dotted) as a function of $m_A$.   For reference, the size of the SD cross section is about $9 \times 10^{-4}$ pb (proton) and $6 \times 10^{-4}$ pb (neutron) and $m_{\chi}=93$ GeV.  The thermal relic density is $\Omega_\mr{DM}\,h^{2} = 0.1$.  The minimum value for the total SI DD is $\sigma_\mr{SI}^\mr{min} = 3\times 10^{-12}$ pb for $m_A = 751$ GeV.  By changing $m_A$ by $5\%$, the cross section becomes $\sim 2\times 10^{-10}$ pb.  For small $m_A$ the cross section is on the order of $\sigma_\mr{SI} \sim 10^{-7}$ pb and in the decoupling limit the cross section is on the order of $\sigma_\mr{SI} \sim 10^{-9}$ pb -- the entire region where there are not any conspiratorial cancellations is within the reach of the next generation of SI experiments.}
\label{fig:SI/SD}
\end{center}
\end{figure}

Here are the options for minimizing $\sigma_\mr{SI}$:
\begin{enumerate}
\item One can make $m_h$ and $m_H$ heavy; however $m_h \approx 115$ GeV in the MSSM in the absence of large fine-tunings.  Even in SplitSUSY, $m_h \lsim 160$ GeV.
\item Since $c_{u,d} \sim (Z_W-t_w\,Z_B)$, \emph{i.e} the Higgs couples to the Zino, one could attempt to restrict the DM to only be a photino-Higgsino admixture.  In Appendix \ref{sec:photinoDM}, we show that this is impossible when one restricts $M_2$ by the LEP bound.
\item One can tune $\left(f^{(N)}_{Tu}+2\,\frac{2}{27}\,f^{(N)}_{TG}\right)\,\frac{c_u}{m_u}$ against $\left( f^{(N)}_{Td} + f^{(N)}_{Ts}+\frac{2}{27}\,f^{(N)}_{TG} \right)\,\frac{c_d}{m_d}$ by tuning the contribution from $H$ against that from $h$.  As we will discuss below, it is not possible to precisely tune this quantity to zero simultaneously for the proton and the neutron (see Fig. \ref{fig:SI/SD}).  However, an approximate realization of this condition is possible -- this is the tuning that underlies large SD/small SI points in Figs. \ref{fig:SIvsSD} and \ref{fig:SIvsSDThermal} and reported in the literature (\emph{e.g.} \cite{MoulinMayet}).
\item One can tune the contribution from the proton against the contribution from the neutron.  The cancellation would only hold for a specific element.  Since all experiments do not use the same elements, we will not pursue this case further.
\end{enumerate}

In what follows, we minimize the SI cross section by tuning the contributions from the $h$ and $H$ against each other (point 3 above).  From Eq.~(\ref{eq:cdMSSM}), this cancellation requires (in the decoupling/large $t_{\beta}$ limit) $\mr{sgn}(Z_{H_u}) = \mr{sgn}(Z_{H_d})$.  This condition for cancellations to be possible was first noted in \cite{Ellis:2000ds}.  Using DarkSUSY we have confirmed that this is a necessary condition, not just in this limit, but for any values of the pseudo-scalar Higgs mass ($m_A$) and $t_{\beta}$.  
This condition only occurs for certain signs of $M_1$, $M_2$ and $\mu$.  If large SD/small SI were observed for neutralino DM, this would constrain the signs in the neutralino mass matrix.  

Let us estimate the maximum allowed suppression.  To good approximation\footnote{From Fig.~\ref{fig:SI/SD} the absolute minimum of the total SI cross section occurs between the region where the coupling to the proton and neutron vanish.  Therefore, the following analytic estimate will be off by a factor of a few.}, the best one can do is to tune away the coupling to (for example) the proton:
\be\label{eq:cancelProtonSI}
\frac{c_u}{m_u} = -\left(\frac{f^{(p)}_{Td}+f^{(p)}_{Ts}+\frac{2}{27}\,f^{(p)}_{TG}}{f^{(p)}_{Tu}+2\,\frac{2}{27}\,f^{(p)}_{TG}}\right) \frac{c_d}{m_d} \equiv - f^{(p)}_{d/u}\,\frac{c_d}{m_d} \approx -1.64\, \frac{c_d}{m_d}.
\ee
In order for Eq.~(\ref{eq:cancelProtonSI}) to have a guaranteed solution requires independent control of $\alpha$ and $m_H$.  Since there is a non-trivial relationship between $\alpha$ and $m_H$ (both are determined by $m_A$), our lower bound provides a conservative estimate.  Using Eq.~(\ref{eq:naievecq}) to estimate $c_q$ and plugging in the relationship between $c_u$ and $c_d$ from Eq.~(\ref{eq:cancelProtonSI}) gives $\sigma_\mr{SI}^p = 0$ and
\bea
\sigma_\mr{SI} &=& \sigma_\mr{SI}^n = \frac{4}{\pi}\,m_n^2\,\frac{(A-Z)^2}{A^2}\,m_r^2\,y_{\chi}^2\,\frac{1}{m_h^4}\nonumber\\
&&\left(\left(f_{Tu}^{(n)}+2\,\frac{2}{27}\,f_{TG}^{(n)}\right)\,f^{(p)}_{d/u}-\left(f_{Td}^{(n)}+f_{Ts}^{(n)}+\frac{2}{27}\,f_{TG}^{(n)}\right)\right)^2\\
&\approx& 8\times 10^{-13}\,\mr{pb}\,\left(\frac{115\,\mr{GeV}}{m_h}\right)^4\,\left(\frac{y_{\chi}}{0.1}\right)^2 \qquad(\textrm{SI with cancellations})\label{eq:largestSISuppression}.
\eea
This gives an estimate for how small SI can be, absent taking some of $M_{1}, M_{2} ,\mu \rightarrow \infty$.  The effects of the current uncertainties on the hadronic matrix elements described in Sec.~\ref{sec:SI} can change the amount of cancellation allowed (the coefficient in Eq.~(\ref{eq:cancelProtonSI})), altering the lower bound in Eq.~(\ref{eq:largestSISuppression}) by $\mathcal{O}(50\%)$.

In Fig. \ref{fig:SI/SD} we show the SI cross section on the proton, the neutron and both as a function of $m_A$ for a 93 GeV neutralino with a thermal relic density of $\Omega_\mr{DM}\,h^{2} = 0.1$, $\sigma_\mr{SD}^p = 9 \times 10^{-4}$ pb and $\sigma_\mr{SD}^n = 6 \times 10^{-4}$ pb.  One can clearly see that both contributions to SI DD cannot both be canceled simultaneously.  At the minimum, $\sigma_\mr{SI}^\mr{min} = 3\times 10^{-12}$ pb for $m_A = 751$ GeV.  For a shift in $m_A$ of $\sim 5\%$, the cross section becomes $\sim 2\times 10^{-10}$ pb -- a change of almost 2 full orders of magnitude.  This emphasizes the delicacy of the cancellation.  Other than in the limited region where the cancellation occurs, the entire range is probeable by the next generation of SI experiments.  

Numerically, we find that for $\sigma_\mr{SD} > 10^{-4}\,\mr{pb}$, the smallest $\sigma_\mr{SI}$ can be is $\mathcal{O}(10^{-14}\,\mr{pb})$ where the suppression beyond the value in Eq.~(\ref{eq:largestSISuppression}) is due to small mixing angles. 

Finally , we note that while these kinds of conspiracies are allowed, there is no reason to expect that the SUSY breaking parameters have anything to do with the nuclear matrix elements.  We take this as evidence that such cancellations are unlikely.

\section{Conclusions}
In this work we have explored the physics of SD DD with an emphasis on the correlations with SI experiments. In the process, we have determined some expectations for the SD cross sections.   In particular, in the MSSM,  $\left(\sigma_\mr{SD}^\mr{SUSY}\right) < 6 \times 10^{-3}\,\mr{pb}$ without making any assumptions about the thermal history.  Again, allowing for a non-trivial cosmic history, but imposing the unified gaugino mass condition, we find $\left(\sigma_\mr{SD}^\mr{SUSY}\right) < 4 \times 10^{-3}\,\mr{pb}$.  Finally,  $\left(\sigma_\mr{SD}^\mr{SUSY}\right) < 2 \times 10^{-3}\,\mr{pb}$ when a  thermal relic density is imposed.  These represent important targets for future experiments.  If one includes the possibility of squark exchange, a SD cross section as high as $2\times 10^{-2}\,\mr{pb}$ can be reached for a neutralino which has a thermal abundance by utilizing the squark pole \cite{Bednyakov:2000he, Bednyakov:2004be}.  We note that in the absence of light squarks, if SD cross sections larger than $\sim 6 \times 10^{-3}\,\mr{pb}$ were observed, the DM would not be an MSSM neutralino.  This would point to more exotic theories like the SDM or models with light mediators \cite{Chang:2009yt}.  For models which reproduce the relic density, in the decoupling limit, and unified gaugino masses, a 1-ton COUPP-like experiment could probe the entire range of SD cross sections up to WIMP masses of $\mathcal{O}(1\,\mr{TeV})$.  

More generally, we have argued that given the experimental constraints from LEP, neutralino DM is likely to be well-tempered with possible signals for the next generation of SI and SD DD experiments.  In fact, any model (such as the SDM) which interacts with the SM via a light Higgs boson can imply a signal in SI experiments and any model of Majorana fermions with non-trivial couplings to the $Z^0$ can imply a signal in SD experiments.  We have enumerated the ways to avoid these arguments.  Since all of these options involve a numerical conspiracy or some new source of tuning, we take them to be disfavored.  With available methods we should be able to probe the majority of the natural range for the SI and SD DD signals of both thermal and non-thermal neutralino DM.

\vspace{-0.2cm}
\subsection*{Acknowledgments}
\vspace{-0.3cm}
\noindent
We thank Daniel Feldman, Gordy Kane, Eric Kuflik, Ilan Levine and Jure Zupan for useful discussions.
The work of T.C. was supported in part by the NSF CAREER Grant NSF-PHY-0743315.
The work of D.J.P. was supported by a Rackham Pre-doctoral Fellowship.
The work of A.P. was supported in part by NSF Career Grant NSF-PHY-0743315 and by DOE Grant \#DE-FG02-95ER40899.

\def\theequation{\Alph{section}.\arabic{equation}}
\begin{appendix}

\setcounter{equation}{0}

\section{Squark Contributions to Direct Detection}\label{sec:SquarkExchange}
The neutralino can scatter off of quarks via $s$-channel squark exchange, giving contributions to $\mathcal{O}^\mr{SI}_q$ or $\mathcal{O}^\mr{SD}_q$.  Only squarks that couple to the light quarks $(u,d,s)$ will be able to contribute to the SI and SD cross sections since only the light quarks have non-negligible nuclear matrix elements.

A non-zero ``left-right" squark mixture is required since SI scattering converts a left-handed quark into a right-handed quark.  Though a Bino/Wino mixture maximizes the coupling between the quarks and the neutralino, the scattering cross section for a pure Bino is of the same order.    

If one makes the standard assumption that left-right squark mixing (\emph{i.e.} $a$-terms) are proportional to Yukawa couplings, then the squark mixing angle is proportional to $m_q/\tilde{m}_q$.  Therefore, all SI couplings will be proportional to a quark mass and there is no enhancement for the light squarks over Higgs boson exchange.  The maximum cross section is 
\be
\left(\sigma_\mr{SI}^\mr{squark}(\chi\,N\rightarrow \chi\,N)\right)_\mr{max} = 6 \times 10^{-9} \textrm{ pb} \left( \frac{\textrm{200 GeV}}{\tilde{m}_s} \right)^4,
\ee
for a Bino-Wino mix.  This is subdominant to the Higgs boson exchange contribution barring the cancellations discussed in Sec.~\ref{sec:SmallSILargeSD}\footnote{If exceptionally large left-right in the squark sector is allowed (perhaps through abnormally large $a$-terms) a contribution to $\sigma_\mr{SI}(\chi\,N\rightarrow \chi\,N)$ of $\mathcal{O}(10^{-3}$ pb) may be obtained.}.

The maximum $\sigma_\mr{SD}(\chi\,p\rightarrow \chi\,p)$ contribution from squark exchange is for a ``left-handed'' up-type squark coupling to a pure Wino, due to the larger $SU(2)$ gauge coupling:
\be
\left(\sigma_\mr{SD}^\mr{squark}(\chi\,p\rightarrow \chi\,p)\right)_\mr{max} = 3 \times 10^{-4} \textrm{ pb} \left( \frac{\textrm{200 GeV}}{\tilde{m}_u} \right)^4.
\ee
This is typically subdominant to the $Z^0$ contribution to SD DD.  Thus, we will focus on the effects of $Z^0$ exchange in our discussions of the expected SD cross section.

\section{The Bino/Higgsino and Wino/Higgsino Limits}\label{sec:BinoAndWinoLimit}
In the limit of large $M_1$ ($M_2$) the neutralino is dominantly a Wino/Higgsino (Bino/Higgsino) admixture.  We can explore this effective 3 state system using the SDM defined as (see Eq.~(\ref{eq:L_SDM}) above):
\begin{equation}
{\mathcal L_\mr{SDM}} \ni \mu_{D}\, D\, \bar{D} + \lambda\, \textit{\textbf{h}}\, S\, D + \lambda^{\prime}\, \textit{\textbf{h}}^{\ast}\, S\, \bar{D} + \frac{\mu_S}{2}\, S^{2}.
\end{equation}
The resulting lightest eigenstate ($\chi$) is specified by
\be
\chi \equiv Z_S\,S + Z_D\,D+Z_{\bar{D}}\,\bar{D}.
\ee
Following \cite{ArkaniHamed:2006mb}, it is useful to write this system in a basis defined by $S$ and $D_{\pm} \equiv\frac{1}{\sqrt{2}}\,(D\pm\,\bar{D})$.  Note that the labels $\pm$ have nothing to do with electric charge.  The mass matrix is then, in the $(S,\,D_+,\,D_-)$ basis,
\[ \mathcal{M}_\mr{SDM} = \left( \begin{array}{ccc}
\mu_S & \frac{1}{\sqrt{2}}\,(\lambda+\lambda')\,v & \frac{1}{\sqrt{2}}\,(\lambda-\lambda')\,v \\
\frac{1}{\sqrt{2}}\,(\lambda+\lambda')\,v & \mu_D & 0 \\
\frac{1}{\sqrt{2}}\,(\lambda-\lambda')\,v & 0 & -\mu_D \end{array} \right),\label{eq:SDMMassMatrix}\]

with the resulting lightest eigenstate,
\be
\chi \equiv Z_S\,S + Z_{D_+}\,D_+ +Z_{D_-}\,D_-.
\ee

Since we are interested in the SD DD cross section, our goal is to extract the coupling of $\chi$ to the $Z^0$.  The coefficient of the operator $\mathcal{O}^\mr{SD}_q$ of Eq.~(\ref{eq:SDOpp}) is given by
\be
d_q = - \frac{g^2}{4\,m_Z^2\,c_w^2}\,|2\,Z_{D_+}\,Z_{D_-}|^2\,T_3^q.
\ee  
Note that $|2\,Z_{D_+}\,Z_{D_-}|\equiv |Z_D|^2-|Z_{\bar{D}}|^2$.
One can find analytic expressions for the mass eigenstates and the combination $|2\,Z_{D_+}\,Z_{D_-}|$ in various useful limits.  To second order in $v$, for $|\mu_D|,\,|\mu_S|,\,(|\mu_D|-|\mu_S|)\gg \lambda\,v,\,\lambda'\,v$ 
\bea
m_{\chi} &=& \mu_S-\frac{2\,\lambda\,\lambda'\,v^2}{\mu_D}-\frac{(\lambda^2+\lambda'^2)\,v^2\,\mu_S}{\mu_D^2}\\
|2\,Z_{D_+}\,Z_{D_-}| &=& \frac{(\lambda'^2-\lambda^2)\,v^2}{\mu_D^2-\mu_S^2},
\eea
and for $|\mu_D| = |\mu_S| \gg \lambda\,v,\,\lambda'\,v$,
\bea
m_{\chi} &=& \mu_S-\frac{1}{\sqrt{2}}\,|\lambda+\lambda'|\,|v|+\frac{(\lambda-\lambda')^2\,v^2}{8\,\mu_S}\\
|2\,Z_{D_+}\,Z_{D_-}| &=& \frac{(\lambda'-\lambda)\,v}{2\,\sqrt{2}\,|\mu_S|}+\frac{(\lambda'^2-\lambda^2)\,v^2}{8\,\mu_S^2}.
\eea
Perturbing away from the limit of exact degeneracy gives corrections to these expressions of $\mathcal{O}((\mu_S-\mu_D)/\mu_D)$.  Note we have assumed that there is no CP violation for simplicity.  In order to apply these expressions to the MSSM one can make the identifications
\begin{center}
  \begin{tabular}{ ||c | c | c || }
    \hline
    \hline
    SDM        & Bino/Higgsino & Wino/Higgsino \\\hline 
    $\mu_S$      & $M_1$ & $M_2$ \\ 
    $\mu_D$      & $\mu$ & $\mu$ \\
    $\lambda\, v$  & $-m_Z \,s_w\,c_{\beta}$ & $m_Z\,c_w\,c_{\beta}$ \\
    $\lambda'\,v$  & $-m_Z\,s_w\,s_{\beta}$ & $m_Z\,c_w\,s_{\beta}$ \\
    \hline
     \hline
  \end{tabular}
\end{center}
where we neglect terms of $\mathcal{O}(1/M_2)$ for the Bino/Higgsino system and $\mathcal{O}(1/M_1)$ for the Wino/Higgsino system.

Explicitly making the substitutions for the MSSM we have

\bea 
|Z_{H_d}|^2-|Z_{H_u}|^2 = \left\{ \begin{array}{ll}
\frac{c_{2\beta}\,s_w^2\,m_Z^2}{\mu^2-M_1^2} & \mbox{for $|M_1|,\,|\mu|,\,|\mu|-|M_1|>m_Z$, $M_2\rightarrow \infty$}\\
\frac{c_{2\beta}\,c_w^2\,m_Z^2}{\mu^2-M_2^2} & \mbox{for $|M_2|,\,|\mu|,\,|\mu|-|M_2|>m_Z$, $M_1\rightarrow \infty$},\end{array} \right. 
\eea
and
\bea 
|Z_{H_d}|^2-|Z_{H_u}|^2 = \left\{ \begin{array}{ll}
\frac{(s_{\beta}-c_{\beta})\,s_w\,m_Z}{2\sqrt{2}\,|\mu|}+\frac{(s_{\beta}^2-c_{\beta}^2)\,s_w^2\,m_Z^2}{8\,\mu^2} & \mbox{for $|M_1| = |\mu| > m_Z$, $M_2\rightarrow \infty$} \\
\frac{(s_{\beta}-c_{\beta})\,c_w\,m_Z}{2\sqrt{2}\,|\mu|}+\frac{(s_{\beta}^2-c_{\beta}^2)\,c_w^2\,m_Z^2}{8\,\mu^2} & \mbox{for $|M_2| = |\mu| > m_Z$, $M_1\rightarrow \infty$}.\end{array} \right.  \eea 

\section{No-go Theorem for photino-Higgsino DM}\label{sec:photinoDM}

The neutralino mass matrix in the $(\tilde{\gamma}, \tilde{Z}, \tilde{H}_d, \tilde{H}_u)$ basis is given by
\[\mathcal{M}=\left( \begin{array}{cccc}
M_1\,c_w^2+ M_2\,s_w^2  & (M_1-M_2)\,c_w\,s_w   & -m_Z\,s_{2w} \, c_{\beta} & m_Z\,s_{2w}\, s_{\beta}\\
(M_1-M_2)\,c_w\,s_w   & M_1\,s_w^2+M_2\,c_w^2 & m_Z\,c_{2w}\, c_{\beta} & -m_Z\,c_{2w}\, s_{\beta}\\
-m_Z\,s_{2w}\, c_{\beta} & m_Z\,c_{2w}\, c_{\beta} & 0 & -\mu  \\
 m_Z\,s_{2w}\, s_{\beta} & -m_Z\,c_{2w}\, s_{\beta} & -\mu & 0  \end{array} \right).\]

Is it possible to generate a large SD/SI ratio by having DM which is only a mixture of photino and Higgsino?  The Higgsino component is required for a non-trivial coupling to the $Z^0$ and an admixture of photino (and not Zino) will allow $(|Z_{H_d}|^2-|Z_{H_u}|^2)\neq 0$ without introducing a coupling to the Higgs.  We show that current phenomenological bounds preclude this possibility.

There are two potential options.  The first is decoupling the Zino by making it heavy while tuning the photino mass to be $\sim \mu$.  This implies taking the limit where $M_1$ and $M_2$ are large while the combination $M_1\,c^2_w+M_2\,s^2_w$ stays small, which requires $\mr{sgn}(M_1)\ne\,\mr{sgn}(M_2)$.  Then the Zino-photino mixing will go like $(M_1-M_2)/m_{\tilde{Z}} > \mathcal{O}(1)$.  Note that we are free to take $M_1 < m_Z$ to suppress this mixing, but due to the LEP bound on the chargino mass, $M_2 > m_Z$.  The second option is to try to eliminate the photino-Zino mixing by taking $M_1 = M_2$.  Then the Zino and photino have the same mass and the Higgsino will mix with both, resulting in a DM state which is an equal admixture of all 4 gauge eigenstates.  Therefore, a neutralino cannot be a mixture of only photino and Higgsino.
\end{appendix}

\bibliography{SIvsSD}{}

\begin{thebibliography}{66}
\expandafter\ifx\csname natexlab\endcsname\relax\def\natexlab#1{#1}\fi
\expandafter\ifx\csname bibnamefont\endcsname\relax
  \def\bibnamefont#1{#1}\fi
\expandafter\ifx\csname bibfnamefont\endcsname\relax
  \def\bibfnamefont#1{#1}\fi
\expandafter\ifx\csname citenamefont\endcsname\relax
  \def\citenamefont#1{#1}\fi
\expandafter\ifx\csname url\endcsname\relax
  \def\url#1{\texttt{#1}}\fi
\expandafter\ifx\csname urlprefix\endcsname\relax\def\urlprefix{URL }\fi
\providecommand{\bibinfo}[2]{#2}
\providecommand{\eprint}[2][]{\url{#2}}

\bibitem[{\citenamefont{Komatsu et~al.}(2009)}]{Komatsu:2008hk}
\bibinfo{author}{\bibfnamefont{E.}~\bibnamefont{Komatsu}} \bibnamefont{et~al.}
  (\bibinfo{collaboration}{WMAP}), \bibinfo{journal}{Astrophys. J. Suppl.}
  \textbf{\bibinfo{volume}{180}}, \bibinfo{pages}{330} (\bibinfo{year}{2009}),
  \eprint{0803.0547}.

\bibitem[{\citenamefont{Adriani et~al.}(2009)}]{PAMELA}
\bibinfo{author}{\bibfnamefont{O.}~\bibnamefont{Adriani}} \bibnamefont{et~al.}
  (\bibinfo{collaboration}{PAMELA}), \bibinfo{journal}{Nature}
  \textbf{\bibinfo{volume}{458}}, \bibinfo{pages}{607} (\bibinfo{year}{2009}),
  \eprint{0810.4995}.

\bibitem[{\citenamefont{Abdo et~al.}(2009)}]{FERMI}
\bibinfo{author}{\bibfnamefont{A.~A.} \bibnamefont{Abdo}} \bibnamefont{et~al.}
  (\bibinfo{collaboration}{The Fermi LAT}), \bibinfo{journal}{Phys. Rev. Lett.}
  \textbf{\bibinfo{volume}{102}}, \bibinfo{pages}{181101}
  (\bibinfo{year}{2009}), \eprint{0905.0025}.

\bibitem[{\citenamefont{Bernabei et~al.}(2000)}]{DAMA}
\bibinfo{author}{\bibfnamefont{R.}~\bibnamefont{Bernabei}} \bibnamefont{et~al.}
  (\bibinfo{collaboration}{DAMA}), \bibinfo{journal}{Phys. Lett.}
  \textbf{\bibinfo{volume}{B480}}, \bibinfo{pages}{23} (\bibinfo{year}{2000}).

\bibitem[{\citenamefont{Bernabei et~al.}(2008)}]{DAMALIBRA}
\bibinfo{author}{\bibfnamefont{R.}~\bibnamefont{Bernabei}} \bibnamefont{et~al.}
  (\bibinfo{collaboration}{DAMA}), \bibinfo{journal}{Eur. Phys. J.}
  \textbf{\bibinfo{volume}{C56}}, \bibinfo{pages}{333} (\bibinfo{year}{2008}),
  \eprint{0804.2741}.

\bibitem[{\citenamefont{Lee and Weinberg}(1977)}]{Lee:1977ua}
\bibinfo{author}{\bibfnamefont{B.~W.} \bibnamefont{Lee}} \bibnamefont{and}
  \bibinfo{author}{\bibfnamefont{S.}~\bibnamefont{Weinberg}},
  \bibinfo{journal}{Phys. Rev. Lett.} \textbf{\bibinfo{volume}{39}},
  \bibinfo{pages}{165} (\bibinfo{year}{1977}).

\bibitem[{\citenamefont{Servant and Tait}(2002)}]{Servant:2002hb}
\bibinfo{author}{\bibfnamefont{G.}~\bibnamefont{Servant}} \bibnamefont{and}
  \bibinfo{author}{\bibfnamefont{T.~M.~P.} \bibnamefont{Tait}},
  \bibinfo{journal}{New J. Phys.} \textbf{\bibinfo{volume}{4}},
  \bibinfo{pages}{99} (\bibinfo{year}{2002}), \eprint{hep-ph/0209262}.

\bibitem[{\citenamefont{Martin}(1997)}]{Martin:1997ns}
\bibinfo{author}{\bibfnamefont{S.~P.} \bibnamefont{Martin}}
  (\bibinfo{year}{1997}), \eprint{hep-ph/9709356}.

\bibitem[{\citenamefont{Arkani-Hamed et~al.}(2006)\citenamefont{Arkani-Hamed,
  Delgado, and Giudice}}]{ArkaniHamed:2006mb}
\bibinfo{author}{\bibfnamefont{N.}~\bibnamefont{Arkani-Hamed}},
  \bibinfo{author}{\bibfnamefont{A.}~\bibnamefont{Delgado}}, \bibnamefont{and}
  \bibinfo{author}{\bibfnamefont{G.~F.} \bibnamefont{Giudice}},
  \bibinfo{journal}{Nucl. Phys.} \textbf{\bibinfo{volume}{B741}},
  \bibinfo{pages}{108} (\bibinfo{year}{2006}), \eprint{hep-ph/0601041}.

\bibitem[{\citenamefont{Moroi and Randall}(2000)}]{Moroi:1999zb}
\bibinfo{author}{\bibfnamefont{T.}~\bibnamefont{Moroi}} \bibnamefont{and}
  \bibinfo{author}{\bibfnamefont{L.}~\bibnamefont{Randall}},
  \bibinfo{journal}{Nucl. Phys.} \textbf{\bibinfo{volume}{B570}},
  \bibinfo{pages}{455} (\bibinfo{year}{2000}), \eprint{hep-ph/9906527}.

\bibitem[{\citenamefont{Catena and Ullio}(2009)}]{Catena:2009mf}
\bibinfo{author}{\bibfnamefont{R.}~\bibnamefont{Catena}} \bibnamefont{and}
  \bibinfo{author}{\bibfnamefont{P.}~\bibnamefont{Ullio}}
  (\bibinfo{year}{2009}), \eprint{0907.0018}.

\bibitem[{\citenamefont{Salucci et~al.}(2010)\citenamefont{Salucci, Nesti,
  Gentile, and Martins}}]{Salucci:2010qr}
\bibinfo{author}{\bibfnamefont{P.}~\bibnamefont{Salucci}},
  \bibinfo{author}{\bibfnamefont{F.}~\bibnamefont{Nesti}},
  \bibinfo{author}{\bibfnamefont{G.}~\bibnamefont{Gentile}}, \bibnamefont{and}
  \bibinfo{author}{\bibfnamefont{C.~F.} \bibnamefont{Martins}}
  (\bibinfo{year}{2010}), \eprint{1003.3101}.

\bibitem[{\citenamefont{Gaitskell}(2004)}]{Gaitskell:2004gd}
\bibinfo{author}{\bibfnamefont{R.~J.} \bibnamefont{Gaitskell}},
  \bibinfo{journal}{Ann. Rev. Nucl. Part. Sci.} \textbf{\bibinfo{volume}{54}},
  \bibinfo{pages}{315} (\bibinfo{year}{2004}).

\bibitem[{\citenamefont{Jungman et~al.}(1996)\citenamefont{Jungman,
  Kamionkowski, and Griest}}]{Jungman:1995df}
\bibinfo{author}{\bibfnamefont{G.}~\bibnamefont{Jungman}},
  \bibinfo{author}{\bibfnamefont{M.}~\bibnamefont{Kamionkowski}},
  \bibnamefont{and} \bibinfo{author}{\bibfnamefont{K.}~\bibnamefont{Griest}},
  \bibinfo{journal}{Phys. Rept.} \textbf{\bibinfo{volume}{267}},
  \bibinfo{pages}{195} (\bibinfo{year}{1996}), \eprint{hep-ph/9506380}.

\bibitem[{\citenamefont{Barger et~al.}(2008)\citenamefont{Barger, Keung, and
  Shaughnessy}}]{Barger:2008qd}
\bibinfo{author}{\bibfnamefont{V.}~\bibnamefont{Barger}},
  \bibinfo{author}{\bibfnamefont{W.-Y.} \bibnamefont{Keung}}, \bibnamefont{and}
  \bibinfo{author}{\bibfnamefont{G.}~\bibnamefont{Shaughnessy}},
  \bibinfo{journal}{Phys. Rev.} \textbf{\bibinfo{volume}{D78}},
  \bibinfo{pages}{056007} (\bibinfo{year}{2008}), \eprint{0806.1962}.

\bibitem[{\citenamefont{Belanger et~al.}(2009)\citenamefont{Belanger, Nezri,
  and Pukhov}}]{Belanger:2008gy}
\bibinfo{author}{\bibfnamefont{G.}~\bibnamefont{Belanger}},
  \bibinfo{author}{\bibfnamefont{E.}~\bibnamefont{Nezri}}, \bibnamefont{and}
  \bibinfo{author}{\bibfnamefont{A.}~\bibnamefont{Pukhov}},
  \bibinfo{journal}{Phys. Rev.} \textbf{\bibinfo{volume}{D79}},
  \bibinfo{pages}{015008} (\bibinfo{year}{2009}), \eprint{0810.1362}.

\bibitem[{\citenamefont{Bertone et~al.}(2007)\citenamefont{Bertone, Cerdeno,
  Collar, and Odom}}]{Bertone:2007xj}
\bibinfo{author}{\bibfnamefont{G.}~\bibnamefont{Bertone}},
  \bibinfo{author}{\bibfnamefont{D.~G.} \bibnamefont{Cerdeno}},
  \bibinfo{author}{\bibfnamefont{J.~I.} \bibnamefont{Collar}},
  \bibnamefont{and} \bibinfo{author}{\bibfnamefont{B.~C.} \bibnamefont{Odom}},
  \bibinfo{journal}{Phys. Rev. Lett.} \textbf{\bibinfo{volume}{99}},
  \bibinfo{pages}{151301} (\bibinfo{year}{2007}), \eprint{0705.2502}.

\bibitem[{\citenamefont{Berger et~al.}(2009)\citenamefont{Berger, Gainer,
  Hewett, and Rizzo}}]{Berger:2008cq}
\bibinfo{author}{\bibfnamefont{C.~F.} \bibnamefont{Berger}},
  \bibinfo{author}{\bibfnamefont{J.~S.} \bibnamefont{Gainer}},
  \bibinfo{author}{\bibfnamefont{J.~L.} \bibnamefont{Hewett}},
  \bibnamefont{and} \bibinfo{author}{\bibfnamefont{T.~G.} \bibnamefont{Rizzo}},
  \bibinfo{journal}{JHEP} \textbf{\bibinfo{volume}{02}}, \bibinfo{pages}{023}
  (\bibinfo{year}{2009}), \eprint{0812.0980}.

\bibitem[{\citenamefont{Ahmed et~al.}(2009{\natexlab{a}})}]{CDMS}
\bibinfo{author}{\bibfnamefont{Z.}~\bibnamefont{Ahmed}} \bibnamefont{et~al.}
  (\bibinfo{collaboration}{CDMS}), \bibinfo{journal}{Phys. Rev. Lett.}
  \textbf{\bibinfo{volume}{102}}, \bibinfo{pages}{011301}
  (\bibinfo{year}{2009}{\natexlab{a}}), \eprint{0802.3530}.

\bibitem[{\citenamefont{Angle et~al.}(2008)}]{XENON}
\bibinfo{author}{\bibfnamefont{J.}~\bibnamefont{Angle}} \bibnamefont{et~al.}
  (\bibinfo{collaboration}{XENON}), \bibinfo{journal}{Phys. Rev. Lett.}
  \textbf{\bibinfo{volume}{100}}, \bibinfo{pages}{021303}
  (\bibinfo{year}{2008}), \eprint{0706.0039}.

\bibitem[{\citenamefont{Ahmed et~al.}(2009{\natexlab{b}})}]{Ahmed:2009zw}
\bibinfo{author}{\bibfnamefont{Z.}~\bibnamefont{Ahmed}} \bibnamefont{et~al.}
  (\bibinfo{collaboration}{The CDMS}) (\bibinfo{year}{2009}{\natexlab{b}}),
  \eprint{0912.3592}.

\bibitem[{\citenamefont{Lee. et~al.}(2007)}]{KIMS}
\bibinfo{author}{\bibfnamefont{H.~S.} \bibnamefont{Lee.}} \bibnamefont{et~al.}
  (\bibinfo{collaboration}{KIMS}), \bibinfo{journal}{Phys. Rev. Lett.}
  \textbf{\bibinfo{volume}{99}}, \bibinfo{pages}{091301}
  (\bibinfo{year}{2007}), \eprint{0704.0423}.

\bibitem[{\citenamefont{Archambault et~al.}(2009)}]{Archambault:2009sm}
\bibinfo{author}{\bibfnamefont{S.}~\bibnamefont{Archambault}}
  \bibnamefont{et~al.}, \bibinfo{journal}{Phys. Lett.}
  \textbf{\bibinfo{volume}{B682}}, \bibinfo{pages}{185} (\bibinfo{year}{2009}),
  \eprint{0907.0307}.

\bibitem[{\citenamefont{Abbasi et~al.}(2009)}]{Abbasi:2009uz}
\bibinfo{author}{\bibfnamefont{R.}~\bibnamefont{Abbasi}} \bibnamefont{et~al.}
  (\bibinfo{collaboration}{ICECUBE}), \bibinfo{journal}{Phys. Rev. Lett.}
  \textbf{\bibinfo{volume}{102}}, \bibinfo{pages}{201302}
  (\bibinfo{year}{2009}), \eprint{0902.2460}.

\bibitem[{\citenamefont{Desai et~al.}(2004)}]{Desai:2004pq}
\bibinfo{author}{\bibfnamefont{S.}~\bibnamefont{Desai}} \bibnamefont{et~al.}
  (\bibinfo{collaboration}{Super-Kamiokande}), \bibinfo{journal}{Phys. Rev.}
  \textbf{\bibinfo{volume}{D70}}, \bibinfo{pages}{083523}
  (\bibinfo{year}{2004}), \eprint{hep-ex/0404025}.

\bibitem[{\citenamefont{Levine}(2009{\natexlab{a}})}]{DPFTalk}
\bibinfo{author}{\bibfnamefont{I.}~\bibnamefont{Levine}},
  \bibinfo{journal}{Talk at DPF}  (\bibinfo{year}{2009}{\natexlab{a}}).

\bibitem[{\citenamefont{Behnke et~al.}(2008)}]{COUPP}
\bibinfo{author}{\bibfnamefont{E.}~\bibnamefont{Behnke}} \bibnamefont{et~al.}
  (\bibinfo{collaboration}{COUPP}), \bibinfo{journal}{Science}
  \textbf{\bibinfo{volume}{319}}, \bibinfo{pages}{933} (\bibinfo{year}{2008}),
  \eprint{0804.2886}.

\bibitem[{\citenamefont{Barnabe-Heider et~al.}(2005)}]{PICASSO}
\bibinfo{author}{\bibfnamefont{M.}~\bibnamefont{Barnabe-Heider}}
  \bibnamefont{et~al.} (\bibinfo{collaboration}{PICASSO}),
  \bibinfo{journal}{Phys. Lett.} \textbf{\bibinfo{volume}{B624}},
  \bibinfo{pages}{186} (\bibinfo{year}{2005}), \eprint{hep-ex/0502028}.

\bibitem[{\citenamefont{Wiebusch et~al.}(2009)}]{Wiebusch:2009jf}
\bibinfo{author}{\bibfnamefont{C.}~\bibnamefont{Wiebusch}} \bibnamefont{et~al.}
  (\bibinfo{collaboration}{ICECUBE}) (\bibinfo{year}{2009}),
  \eprint{0907.2263}.

\bibitem[{\citenamefont{Levine}(2009{\natexlab{b}})}]{LevineConversation}
\bibinfo{author}{\bibfnamefont{I.}~\bibnamefont{Levine}},
  \bibinfo{journal}{Private Communication}
  (\bibinfo{year}{2009}{\natexlab{b}}).

\bibitem[{\citenamefont{Shifman et~al.}(1978)\citenamefont{Shifman, Vainshtein,
  and Zakharov}}]{Shifman:1978zn}
\bibinfo{author}{\bibfnamefont{M.~A.} \bibnamefont{Shifman}},
  \bibinfo{author}{\bibfnamefont{A.~I.} \bibnamefont{Vainshtein}},
  \bibnamefont{and} \bibinfo{author}{\bibfnamefont{V.~I.}
  \bibnamefont{Zakharov}}, \bibinfo{journal}{Phys. Lett.}
  \textbf{\bibinfo{volume}{B78}}, \bibinfo{pages}{443} (\bibinfo{year}{1978}).

\bibitem[{\citenamefont{Freedman et~al.}(1977)\citenamefont{Freedman, Schramm,
  and Tubbs}}]{Freedman:1977xn}
\bibinfo{author}{\bibfnamefont{D.~Z.} \bibnamefont{Freedman}},
  \bibinfo{author}{\bibfnamefont{D.~N.} \bibnamefont{Schramm}},
  \bibnamefont{and} \bibinfo{author}{\bibfnamefont{D.~L.} \bibnamefont{Tubbs}},
  \bibinfo{journal}{Ann. Rev. Nucl. Part. Sci.} \textbf{\bibinfo{volume}{27}},
  \bibinfo{pages}{167} (\bibinfo{year}{1977}).

\bibitem[{\citenamefont{Gondolo et~al.}(2004)}]{DarkSUSY}
\bibinfo{author}{\bibfnamefont{P.}~\bibnamefont{Gondolo}} \bibnamefont{et~al.},
  \bibinfo{journal}{JCAP} \textbf{\bibinfo{volume}{0407}}, \bibinfo{pages}{008}
  (\bibinfo{year}{2004}), \eprint{astro-ph/0406204}.

\bibitem[{\citenamefont{Giedt et~al.}(2009)\citenamefont{Giedt, Thomas, and
  Young}}]{Giedt:2009mr}
\bibinfo{author}{\bibfnamefont{J.}~\bibnamefont{Giedt}},
  \bibinfo{author}{\bibfnamefont{A.~W.} \bibnamefont{Thomas}},
  \bibnamefont{and} \bibinfo{author}{\bibfnamefont{R.~D.} \bibnamefont{Young}}
  (\bibinfo{year}{2009}), \eprint{0907.4177}.

\bibitem[{\citenamefont{Ellis et~al.}(2008)\citenamefont{Ellis, Olive, and
  Savage}}]{EllisOliveSavage}
\bibinfo{author}{\bibfnamefont{J.~R.} \bibnamefont{Ellis}},
  \bibinfo{author}{\bibfnamefont{K.~A.} \bibnamefont{Olive}}, \bibnamefont{and}
  \bibinfo{author}{\bibfnamefont{C.}~\bibnamefont{Savage}},
  \bibinfo{journal}{Phys. Rev.} \textbf{\bibinfo{volume}{D77}},
  \bibinfo{pages}{065026} (\bibinfo{year}{2008}), \eprint{0801.3656}.

\bibitem[{\citenamefont{Barbieri et~al.}(1989)\citenamefont{Barbieri, Frigeni,
  and Giudice}}]{Barbieri:1988zs}
\bibinfo{author}{\bibfnamefont{R.}~\bibnamefont{Barbieri}},
  \bibinfo{author}{\bibfnamefont{M.}~\bibnamefont{Frigeni}}, \bibnamefont{and}
  \bibinfo{author}{\bibfnamefont{G.~F.} \bibnamefont{Giudice}},
  \bibinfo{journal}{Nucl. Phys.} \textbf{\bibinfo{volume}{B313}},
  \bibinfo{pages}{725} (\bibinfo{year}{1989}).

\bibitem[{\citenamefont{Ellis et~al.}(2001)\citenamefont{Ellis, Ferstl, and
  Olive}}]{Ellis:2000jd}
\bibinfo{author}{\bibfnamefont{J.~R.} \bibnamefont{Ellis}},
  \bibinfo{author}{\bibfnamefont{A.}~\bibnamefont{Ferstl}}, \bibnamefont{and}
  \bibinfo{author}{\bibfnamefont{K.~A.} \bibnamefont{Olive}},
  \bibinfo{journal}{Phys. Rev.} \textbf{\bibinfo{volume}{D63}},
  \bibinfo{pages}{065016} (\bibinfo{year}{2001}), \eprint{hep-ph/0007113}.

\bibitem[{\citenamefont{Baer et~al.}(2007)\citenamefont{Baer, Mustafayev, Park,
  and Tata}}]{Baer:2006te}
\bibinfo{author}{\bibfnamefont{H.}~\bibnamefont{Baer}},
  \bibinfo{author}{\bibfnamefont{A.}~\bibnamefont{Mustafayev}},
  \bibinfo{author}{\bibfnamefont{E.-K.} \bibnamefont{Park}}, \bibnamefont{and}
  \bibinfo{author}{\bibfnamefont{X.}~\bibnamefont{Tata}},
  \bibinfo{journal}{JCAP} \textbf{\bibinfo{volume}{0701}}, \bibinfo{pages}{017}
  (\bibinfo{year}{2007}), \eprint{hep-ph/0611387}.

\bibitem[{\citenamefont{Hisano et~al.}(2009)\citenamefont{Hisano, Nakayama, and
  Yamanaka}}]{Hisano:2009xv}
\bibinfo{author}{\bibfnamefont{J.}~\bibnamefont{Hisano}},
  \bibinfo{author}{\bibfnamefont{K.}~\bibnamefont{Nakayama}}, \bibnamefont{and}
  \bibinfo{author}{\bibfnamefont{M.}~\bibnamefont{Yamanaka}}
  (\bibinfo{year}{2009}), \eprint{0912.4701}.

\bibitem[{\citenamefont{Griest and Seckel}(1991)}]{Griest:1990kh}
\bibinfo{author}{\bibfnamefont{K.}~\bibnamefont{Griest}} \bibnamefont{and}
  \bibinfo{author}{\bibfnamefont{D.}~\bibnamefont{Seckel}},
  \bibinfo{journal}{Phys. Rev.} \textbf{\bibinfo{volume}{D43}},
  \bibinfo{pages}{3191} (\bibinfo{year}{1991}).

\bibitem[{\citenamefont{Ellis et~al.}(1998)\citenamefont{Ellis, Falk, and
  Olive}}]{Ellis:1998kh}
\bibinfo{author}{\bibfnamefont{J.~R.} \bibnamefont{Ellis}},
  \bibinfo{author}{\bibfnamefont{T.}~\bibnamefont{Falk}}, \bibnamefont{and}
  \bibinfo{author}{\bibfnamefont{K.~A.} \bibnamefont{Olive}},
  \bibinfo{journal}{Phys. Lett.} \textbf{\bibinfo{volume}{B444}},
  \bibinfo{pages}{367} (\bibinfo{year}{1998}), \eprint{hep-ph/9810360}.

\bibitem[{\citenamefont{Ellis et~al.}(2000{\natexlab{a}})\citenamefont{Ellis,
  Falk, Olive, and Srednicki}}]{Ellis:1999mm}
\bibinfo{author}{\bibfnamefont{J.~R.} \bibnamefont{Ellis}},
  \bibinfo{author}{\bibfnamefont{T.}~\bibnamefont{Falk}},
  \bibinfo{author}{\bibfnamefont{K.~A.} \bibnamefont{Olive}}, \bibnamefont{and}
  \bibinfo{author}{\bibfnamefont{M.}~\bibnamefont{Srednicki}},
  \bibinfo{journal}{Astropart. Phys.} \textbf{\bibinfo{volume}{13}},
  \bibinfo{pages}{181} (\bibinfo{year}{2000}{\natexlab{a}}),
  \eprint{hep-ph/9905481}.

\bibitem[{\citenamefont{Drees and Nojiri}(1993)}]{Drees:1992am}
\bibinfo{author}{\bibfnamefont{M.}~\bibnamefont{Drees}} \bibnamefont{and}
  \bibinfo{author}{\bibfnamefont{M.~M.} \bibnamefont{Nojiri}},
  \bibinfo{journal}{Phys. Rev.} \textbf{\bibinfo{volume}{D47}},
  \bibinfo{pages}{376} (\bibinfo{year}{1993}), \eprint{hep-ph/9207234}.

\bibitem[{\citenamefont{Roszkowski et~al.}(2001)\citenamefont{Roszkowski,
  Ruiz~de Austri, and Nihei}}]{Roszkowski:2001sb}
\bibinfo{author}{\bibfnamefont{L.}~\bibnamefont{Roszkowski}},
  \bibinfo{author}{\bibfnamefont{R.}~\bibnamefont{Ruiz~de Austri}},
  \bibnamefont{and} \bibinfo{author}{\bibfnamefont{T.}~\bibnamefont{Nihei}},
  \bibinfo{journal}{JHEP} \textbf{\bibinfo{volume}{08}}, \bibinfo{pages}{024}
  (\bibinfo{year}{2001}), \eprint{hep-ph/0106334}.

\bibitem[{\citenamefont{Djouadi et~al.}(2005)\citenamefont{Djouadi, Drees, and
  Kneur}}]{Djouadi:2005dz}
\bibinfo{author}{\bibfnamefont{A.}~\bibnamefont{Djouadi}},
  \bibinfo{author}{\bibfnamefont{M.}~\bibnamefont{Drees}}, \bibnamefont{and}
  \bibinfo{author}{\bibfnamefont{J.-L.} \bibnamefont{Kneur}},
  \bibinfo{journal}{Phys. Lett.} \textbf{\bibinfo{volume}{B624}},
  \bibinfo{pages}{60} (\bibinfo{year}{2005}), \eprint{hep-ph/0504090}.

\bibitem[{\citenamefont{Nath and Arnowitt}(1993)}]{Nath:1992ty}
\bibinfo{author}{\bibfnamefont{P.}~\bibnamefont{Nath}} \bibnamefont{and}
  \bibinfo{author}{\bibfnamefont{R.~L.} \bibnamefont{Arnowitt}},
  \bibinfo{journal}{Phys. Rev. Lett.} \textbf{\bibinfo{volume}{70}},
  \bibinfo{pages}{3696} (\bibinfo{year}{1993}), \eprint{hep-ph/9302318}.

\bibitem[{\citenamefont{Feng et~al.}(2001)\citenamefont{Feng, Matchev, and
  Wilczek}}]{Feng:2000zu}
\bibinfo{author}{\bibfnamefont{J.~L.} \bibnamefont{Feng}},
  \bibinfo{author}{\bibfnamefont{K.~T.} \bibnamefont{Matchev}},
  \bibnamefont{and} \bibinfo{author}{\bibfnamefont{F.}~\bibnamefont{Wilczek}},
  \bibinfo{journal}{Phys. Rev.} \textbf{\bibinfo{volume}{D63}},
  \bibinfo{pages}{045024} (\bibinfo{year}{2001}), \eprint{astro-ph/0008115}.

\bibitem[{\citenamefont{Feng et~al.}(2000)\citenamefont{Feng, Matchev, and
  Wilczek}}]{Feng:2000gh}
\bibinfo{author}{\bibfnamefont{J.~L.} \bibnamefont{Feng}},
  \bibinfo{author}{\bibfnamefont{K.~T.} \bibnamefont{Matchev}},
  \bibnamefont{and} \bibinfo{author}{\bibfnamefont{F.}~\bibnamefont{Wilczek}},
  \bibinfo{journal}{Phys. Lett.} \textbf{\bibinfo{volume}{B482}},
  \bibinfo{pages}{388} (\bibinfo{year}{2000}), \eprint{hep-ph/0004043}.

\bibitem[{\citenamefont{Kitano and Nomura}(2005)}]{Kitano:2005wc}
\bibinfo{author}{\bibfnamefont{R.}~\bibnamefont{Kitano}} \bibnamefont{and}
  \bibinfo{author}{\bibfnamefont{Y.}~\bibnamefont{Nomura}},
  \bibinfo{journal}{Phys. Lett.} \textbf{\bibinfo{volume}{B631}},
  \bibinfo{pages}{58} (\bibinfo{year}{2005}), \eprint{hep-ph/0509039}.

\bibitem[{\citenamefont{Acharya et~al.}(2009)\citenamefont{Acharya, Kane,
  Watson, and Kumar}}]{Acharya:2009zt}
\bibinfo{author}{\bibfnamefont{B.~S.} \bibnamefont{Acharya}},
  \bibinfo{author}{\bibfnamefont{G.}~\bibnamefont{Kane}},
  \bibinfo{author}{\bibfnamefont{S.}~\bibnamefont{Watson}}, \bibnamefont{and}
  \bibinfo{author}{\bibfnamefont{P.}~\bibnamefont{Kumar}},
  \bibinfo{journal}{Phys. Rev.} \textbf{\bibinfo{volume}{D80}},
  \bibinfo{pages}{083529} (\bibinfo{year}{2009}), \eprint{0908.2430}.

\bibitem[{\citenamefont{Kamionkowski and Turner}(1990)}]{Kamionkowski:1990ni}
\bibinfo{author}{\bibfnamefont{M.}~\bibnamefont{Kamionkowski}}
  \bibnamefont{and} \bibinfo{author}{\bibfnamefont{M.~S.}
  \bibnamefont{Turner}}, \bibinfo{journal}{Phys. Rev.}
  \textbf{\bibinfo{volume}{D42}}, \bibinfo{pages}{3310} (\bibinfo{year}{1990}).

\bibitem[{\citenamefont{Hall et~al.}(2009)\citenamefont{Hall, Jedamzik,
  March-Russell, and West}}]{Hall:2009bx}
\bibinfo{author}{\bibfnamefont{L.~J.} \bibnamefont{Hall}},
  \bibinfo{author}{\bibfnamefont{K.}~\bibnamefont{Jedamzik}},
  \bibinfo{author}{\bibfnamefont{J.}~\bibnamefont{March-Russell}},
  \bibnamefont{and} \bibinfo{author}{\bibfnamefont{S.~M.} \bibnamefont{West}}
  (\bibinfo{year}{2009}), \eprint{0911.1120}.

\bibitem[{\citenamefont{Kaplan et~al.}(2009)\citenamefont{Kaplan, Luty, and
  Zurek}}]{Kaplan:2009ag}
\bibinfo{author}{\bibfnamefont{D.~E.} \bibnamefont{Kaplan}},
  \bibinfo{author}{\bibfnamefont{M.~A.} \bibnamefont{Luty}}, \bibnamefont{and}
  \bibinfo{author}{\bibfnamefont{K.~M.} \bibnamefont{Zurek}},
  \bibinfo{journal}{Phys. Rev.} \textbf{\bibinfo{volume}{D79}},
  \bibinfo{pages}{115016} (\bibinfo{year}{2009}), \eprint{0901.4117}.

\bibitem[{\citenamefont{Gelmini and Gondolo}(2006)}]{Gelmini:2006pw}
\bibinfo{author}{\bibfnamefont{G.~B.} \bibnamefont{Gelmini}} \bibnamefont{and}
  \bibinfo{author}{\bibfnamefont{P.}~\bibnamefont{Gondolo}},
  \bibinfo{journal}{Phys. Rev.} \textbf{\bibinfo{volume}{D74}},
  \bibinfo{pages}{023510} (\bibinfo{year}{2006}), \eprint{hep-ph/0602230}.

\bibitem[{\citenamefont{Cohen et~al.}(2008)\citenamefont{Cohen, Morrissey, and
  Pierce}}]{Cohen:2008nb}
\bibinfo{author}{\bibfnamefont{T.}~\bibnamefont{Cohen}},
  \bibinfo{author}{\bibfnamefont{D.~E.} \bibnamefont{Morrissey}},
  \bibnamefont{and} \bibinfo{author}{\bibfnamefont{A.}~\bibnamefont{Pierce}},
  \bibinfo{journal}{Phys. Rev.} \textbf{\bibinfo{volume}{D78}},
  \bibinfo{pages}{111701} (\bibinfo{year}{2008}), \eprint{0808.3994}.

\bibitem[{\citenamefont{Arkani-Hamed et~al.}(2005)\citenamefont{Arkani-Hamed,
  Dimopoulos, and Kachru}}]{ArkaniHamedFN}
\bibinfo{author}{\bibfnamefont{N.}~\bibnamefont{Arkani-Hamed}},
  \bibinfo{author}{\bibfnamefont{S.}~\bibnamefont{Dimopoulos}},
  \bibnamefont{and} \bibinfo{author}{\bibfnamefont{S.}~\bibnamefont{Kachru}}
  (\bibinfo{year}{2005}), \eprint{hep-th/0501082}.

\bibitem[{\citenamefont{Mahbubani and Senatore}(2006)}]{SenatoreMahbubani}
\bibinfo{author}{\bibfnamefont{R.}~\bibnamefont{Mahbubani}} \bibnamefont{and}
  \bibinfo{author}{\bibfnamefont{L.}~\bibnamefont{Senatore}},
  \bibinfo{journal}{Phys. Rev.} \textbf{\bibinfo{volume}{D73}},
  \bibinfo{pages}{043510} (\bibinfo{year}{2006}), \eprint{hep-ph/0510064}.

\bibitem[{\citenamefont{Peskin and Wells}(2001)}]{PeskinWells}
\bibinfo{author}{\bibfnamefont{M.~E.} \bibnamefont{Peskin}} \bibnamefont{and}
  \bibinfo{author}{\bibfnamefont{J.~D.} \bibnamefont{Wells}},
  \bibinfo{journal}{Phys. Rev.} \textbf{\bibinfo{volume}{D64}},
  \bibinfo{pages}{093003} (\bibinfo{year}{2001}), \eprint{hep-ph/0101342}.

\bibitem[{\citenamefont{Mandic et~al.}(2000)\citenamefont{Mandic, Pierce,
  Gondolo, and Murayama}}]{Mandic:2000jz}
\bibinfo{author}{\bibfnamefont{V.}~\bibnamefont{Mandic}},
  \bibinfo{author}{\bibfnamefont{A.}~\bibnamefont{Pierce}},
  \bibinfo{author}{\bibfnamefont{P.}~\bibnamefont{Gondolo}}, \bibnamefont{and}
  \bibinfo{author}{\bibfnamefont{H.}~\bibnamefont{Murayama}}
  (\bibinfo{year}{2000}), \eprint{hep-ph/0008022}.

\bibitem[{\citenamefont{Kitano and Nomura}(2006)}]{Kitano:2005ew}
\bibinfo{author}{\bibfnamefont{R.}~\bibnamefont{Kitano}} \bibnamefont{and}
  \bibinfo{author}{\bibfnamefont{Y.}~\bibnamefont{Nomura}},
  \bibinfo{journal}{Phys. Lett.} \textbf{\bibinfo{volume}{B632}},
  \bibinfo{pages}{162} (\bibinfo{year}{2006}), \eprint{hep-ph/0509221}.

\bibitem[{\citenamefont{Hisano et~al.}(2005)\citenamefont{Hisano, Matsumoto,
  Nojiri, and Saito}}]{Hisano:2004pv}
\bibinfo{author}{\bibfnamefont{J.}~\bibnamefont{Hisano}},
  \bibinfo{author}{\bibfnamefont{S.}~\bibnamefont{Matsumoto}},
  \bibinfo{author}{\bibfnamefont{M.~M.} \bibnamefont{Nojiri}},
  \bibnamefont{and} \bibinfo{author}{\bibfnamefont{O.}~\bibnamefont{Saito}},
  \bibinfo{journal}{Phys. Rev.} \textbf{\bibinfo{volume}{D71}},
  \bibinfo{pages}{015007} (\bibinfo{year}{2005}), \eprint{hep-ph/0407168}.

\bibitem[{\citenamefont{Moulin et~al.}(2005)\citenamefont{Moulin, Mayet, and
  Santos}}]{MoulinMayet}
\bibinfo{author}{\bibfnamefont{E.}~\bibnamefont{Moulin}},
  \bibinfo{author}{\bibfnamefont{F.}~\bibnamefont{Mayet}}, \bibnamefont{and}
  \bibinfo{author}{\bibfnamefont{D.}~\bibnamefont{Santos}},
  \bibinfo{journal}{Phys. Lett.} \textbf{\bibinfo{volume}{B614}},
  \bibinfo{pages}{143} (\bibinfo{year}{2005}), \eprint{astro-ph/0503436}.

\bibitem[{\citenamefont{Ellis et~al.}(2000{\natexlab{b}})\citenamefont{Ellis,
  Ferstl, and Olive}}]{Ellis:2000ds}
\bibinfo{author}{\bibfnamefont{J.~R.} \bibnamefont{Ellis}},
  \bibinfo{author}{\bibfnamefont{A.}~\bibnamefont{Ferstl}}, \bibnamefont{and}
  \bibinfo{author}{\bibfnamefont{K.~A.} \bibnamefont{Olive}},
  \bibinfo{journal}{Phys. Lett.} \textbf{\bibinfo{volume}{B481}},
  \bibinfo{pages}{304} (\bibinfo{year}{2000}{\natexlab{b}}),
  \eprint{hep-ph/0001005}.

\bibitem[{\citenamefont{Bednyakov and
  Klapdor-Kleingrothaus}(2001)}]{Bednyakov:2000he}
\bibinfo{author}{\bibfnamefont{V.~A.} \bibnamefont{Bednyakov}}
  \bibnamefont{and} \bibinfo{author}{\bibfnamefont{H.~V.}
  \bibnamefont{Klapdor-Kleingrothaus}}, \bibinfo{journal}{Phys. Rev.}
  \textbf{\bibinfo{volume}{D63}}, \bibinfo{pages}{095005}
  (\bibinfo{year}{2001}), \eprint{hep-ph/0011233}.

\bibitem[{\citenamefont{Bednyakov and
  Klapdor-Kleingrothaus}(2004)}]{Bednyakov:2004be}
\bibinfo{author}{\bibfnamefont{V.~A.} \bibnamefont{Bednyakov}}
  \bibnamefont{and} \bibinfo{author}{\bibfnamefont{H.~V.}
  \bibnamefont{Klapdor-Kleingrothaus}}, \bibinfo{journal}{Phys. Rev.}
  \textbf{\bibinfo{volume}{D70}}, \bibinfo{pages}{096006}
  (\bibinfo{year}{2004}), \eprint{hep-ph/0404102}.

\bibitem[{\citenamefont{Chang et~al.}(2009)\citenamefont{Chang, Pierce, and
  Weiner}}]{Chang:2009yt}
\bibinfo{author}{\bibfnamefont{S.}~\bibnamefont{Chang}},
  \bibinfo{author}{\bibfnamefont{A.}~\bibnamefont{Pierce}}, \bibnamefont{and}
  \bibinfo{author}{\bibfnamefont{N.}~\bibnamefont{Weiner}}
  (\bibinfo{year}{2009}), \eprint{0908.3192}.

\end{thebibliography}

\end{document}